\documentstyle[aps,prl,multicol,epsf,epsfig]{revtex}

\newcommand{\beq}{\begin{equation}}

\newcommand{\eeq}{\end{equation}}
\newcommand{\barray}{\begin{eqnarray}}
\newcommand{\earray}{\end{eqnarray}}
\newcommand{\invol}{\frac{1}{{\cal L}}}
\newcommand{\V}[1]{V_{{#1}}}
\newcommand{\ve}[1]{\vec{#1}}
\newcommand{\ven}{\vec{\eta}}
\newcommand{\venp}{\vec{\eta'}}
\newcommand{\venpp}{\vec{\eta''}}
\newcommand{\ha}[1]{\hat{#1}}

\newcommand{\jq}{\hat{J}^Q_x}
\newcommand{\je}{\hat{J}^E_x}
\newcommand{\vx}[1]{v^x_{{#1}}}
\newcommand{\jc}{\hat{J}_x}
\newcommand{\Grad}{\nabla}

\newcommand{\ep}{\varepsilon_{\ve{p}}}
\newcommand{\ch}{\tilde{c}}
\newcommand{\chd}{\tilde{c}^\dagger}

\begin{document}
\title{  A Sum  Rule for Thermal Conductivity   and Dynamical Thermal Transport Coefficients in  Condensed  Matter -I}
\author{ B Sriram Shastry }
\address{Physics Department, University of California,  Santa Cruz, Ca 95064 }
\date{\today}
\maketitle
\begin{abstract}
We display an interesting sum rule for the dynamical thermal conductivity for many standard  models of condensed matter in terms of the expectation of  a thermal operator. We present the thermal operator for several model systems of current interest,  which enable an evaluation of the sum rule and the  Lorentz number, the thermo electric figure of merit  as well as the thermopower at high frequencies. As a by product,  we present exact formulae for the $T=0$  chemical potential  $\mu(0)$  for charged many-body systems, including the Hubbard model, in terms of expectation values  of extensive operators.
Simple estimates are provided for the thermopower of an infinitely correlated band model on the triangular lattice, modeling the physics of the sodium cobalt oxide system. The present result  goes beyond  the Heikes Mott formula for the thermopower, and   contains an additional transport correction that is sensitive to the lattice topology as well as the sign of hopping.
\end{abstract}
\maketitle

\section{Introduction}

A problem of considerable current interest is  the evaluation of the thermal response functions; $\kappa$ the thermal conductivity and 
$S$ the thermoelectric response function (Seebeck coefficient) for complex systems. These include strongly correlated matter, where the role of interactions is profound and impossible to capture within perturbative formulations, and also materials with complex unit cells, such as the skutterudites\cite{skutterudites}, and negative thermal expansion systems where geometrical frustration plays a major role\cite{nte}. One major motivation  for these efforts is from the requirements of  the thermoelectric industry, where a large thermoelectric  figure of merit  (defined in Eq(\ref{powerfactor})) is desirable , and it is important to understand the material characteristics that promote a large figure of merit.
The standard tool, namely the Boltzmann equation approach, is overly restrictive in that it  is tied down to the concept of long lived and weakly interacting quasi particles, which are  less useful in these contexts. 

In this paper, we present a new formalism for computation of  the thermal response functions, by considering the response to {\em dynamical } temperature gradients, i.e. a frequency dependent problem. This yields the frequency dependent thermal conductivity $\kappa(\omega)$  as well as the thermopower $S(\omega)$. The frequency dependence of $\kappa(\omega)$ is of relevance theoretically and also practically:   a silicon chip  operating at  several gigahertz clock speed in a computing  processor needs to 
transport the { AC} Joule dissipation.  The relevant transport parameter in this context is the dynamical $\kappa(\omega)$,  which can be much smaller    than the  static  $\kappa$ \cite{volz} for $\omega \sim 100$GHz. 

Although  studying   the dynamical conductivity  looks at first sight like further  complicating an already intractable problem, in fact it turns out that several simplifications arise in the limit of high frequencies. These  have a close counterpart in our fairly successful theory of the Hall constant of strongly correlated matter\cite{ssshall}.  In brief, the problem of transport
may be decomposed into the following conceptually distinct  but technically intertwined aspects (a) transport proper (b) single particle description and (c) interactions. The single particle aspect determines such variables  as  the density of states and its derivatives, the transport aspect involves the elastic and inelastic scattering rates usually modeled by a relaxation time, and the interactions aspect involves the influence of correlations. While none of these are trivial, the last one, namely interactions is conceptually as well as computationally most difficult. For example in Mott Hubbard systems, the very notion of the hole is non trivial in the context of say the Hall constant, where one knows that near half filling, the momentum space definition can give opposite results in certain cases to the real space picture for the sign of the Hall constant\cite{ssshall}. 

In this context, going to high frequencies is a great advantage if one can isolate combinations of thermal response functions that are relatively less frequency dependent. In the Hall constant problem, the Hall resistivity $\rho^{xy}$ is quite  benign, being $\omega$ independent at least in the Drude theory, as opposed to the Hall conductivities $ \sigma^{xy}$ and $ \sigma^{xx}$, which are   serious functions of $\omega$ {\em individually} but {\em not} in the combination leading to $\rho^{xy}$. Moreover the high frequency limit is the starting point of the Mori type
treatment of transport quantities, as seen in the example for the Hall constant\cite{lange}. In a similar spirit, we identify several variables in this work, and  write down the operators that need to be evaluated. We stress that the expectation values of the objects computed here are equilibrium values, and while quite non trivial in detail, are conceptually much easier than the transport objects- this is the advantage of disentangling dynamics from interactions as in other contexts. While our approach is not exact either, as it depends on the assumption of weak frequency dependence of such combinations, it is quite distinct from the Boltzmann approach and provides a useful counterpoint to the latter, treating the effect of interactions more respectfully. In simple examples, we show that the weak $\omega$ dependence is an excellent assumption, and further our
results  should provide a stimulus for experimental investigation of the frequency dependence of thermal transport constants. 

We present an interesting sum rule for the real part of the dynamical thermal conductivity. The sum rule is in terms of the expectation of an operator that is presented for several models of condensed matter physics, and while 
the task of evaluating it is non trivial in detail, it is quite feasible to estimate it from other considerations that we present. The situation is akin to the well known plasma or the f-sum rule; the Thomas Ritchie Kuhn sum rule gives the integral of the real conductivity     in terms of easily measurable variables in the continuum $4 \pi n q_e^2 /m$,  whereas {\em on a lattice}, it is (see  Eq(\ref{fsumrule})) the expectation  of essentially the kinetic energy  \cite{plasmasumrule}. 
Such an expectation is non trivial, but several numerical techniques lend themselves to this task quite effectively, such as exact diagonalization and Quantum Monte Carlo methods. This new sum rule should be of considerable interest in many situations, including the case of non linear lattices, where the classical counterpart is already interesting, since it is the expectation of a single extensive operator,  and apparently equally unknown.

Our paper has a twofold purpose, we present the formalism for the thermal transport variables in some detail with the hope of returning to calculate some of them in a later work, and the other is somewhat pedagogical in nature. Existing literature is quite focussed on specific issues, and many subtleties receive less mention than their due.
 It is my impression  that a  detailed  description of the formalism might be useful to  some  workers.  The results of computations based on the formalism presented here should shed light on currently interesting issues, such as the departure from Wiedemann- Franz ratio for the case of strongly correlated systems as seen in experiments, where the Lorentz number departs from the free electron value at low $T$\cite{lorentz}.

An interesting  by-product of our work is the exact formula for the chemical potential $\mu(T)$ at $T=0$ for charged  many body systems in terms of the expectation of an extensive operator. 
In general the only  way of computing $\mu(T)$,  including in the ground state, is from the thermodynamic formula  $\mu(T) = \frac{\partial F(T,N)}{\partial N}$. However  at $T=0$, our result shows an alternate and potentially widely useful method. This arises from the physical requirement  of the vanishing of thermopower {\em at all frequencies} in the ground state, being related to entropy transport, and since the  high frequency formulae explicitly involve the chemical potential,   these yield the new  formulae  for the zero temperature chemical potential of all charged many body systems. We comment on this further in the discussion section (X) .
The plan of the paper is as follows.
\begin{itemize}
\item  In Sections II and III, we derive the complete expressions for the finite frequency thermal conductivity and thermopower, using standard linear response theory, and obtain a  sum rule for the dynamical thermal conductivity. It is expressed in terms of a new extensive object $\Theta^{xx}$ that we term as the ``thermal operator''.
 
\item Section IV makes a connection with the ``standard'' Kubo formulas. Our results contain corrections to the Kubo formulas for the case of non-dissipative systems such as superconductors, and it might be of interest  to see the origin of these in the fine tuning of the standard derivations, as provided in Section IV.
\item In Section V, we identify the  high frequency 
observables that are suitable for study being  less sensitive to $\omega$.  The high frequency Seebeck coefficient is amongst these, and is expressed in terms of another new extensive object $\Phi^{xx}$ that we  denote as the ``thermoelectric operator''.   
\item  The thermal and thermoelectric operators have a dependence on the interactions as well the details of the specific model systems. Hence it is important to compute these operators for standard models, which is accomplished in the latter sections. In Section VI, we present the thermal operator $\Theta^{xx}$ for the anharmonic disordered lattice vibration problem, and make a connection with the classical limit. Illustrative examples from harmonic lattice  are given to provide a feel for the nature of the operators encountered.
 \item In Sec VII, we study the intermediate coupling models of current interest, namely the Hubbard model, the homogeneous electron gas and the periodic Anderson lattice, for which we present the thermal and thermoelectric operators, and also indicate the role of disorder in modifying these variables. We establish the essential correctness of this approach for the case of zero coupling, where the standard Boltzmann- Drude results for the Lorentz number and the thermopower are  reproduced. 
\item In Sec VIII, we study the thermal and thermoelectric operators for the strong coupling models:   the Heisenberg model
and the infinite U Hubbard model.
\item In Sec IX we apply the results of Sec VIII to the currently interesting case of  the triangular lattice $Na_x Co O_2$, a metallic thermoelectric material. Our theory yields a formula for the High T limit of the Seebeck coefficient for  this system, which contains the Heikes Mott formula and provides a transport correction to the same.  The large transport correction highlights the role of the lattice topology.
\item In Sec X, we discuss the special nature of the thermal  type operators, namely the vanishing of its expectation in the ground state. We also comment on the exact formulae for the chemical potential at $T=0$ for charged fermi systems that our approach yields. 
\item  Appendix A contains the already well known results for the electrical conductivity for completeness.  
\end{itemize}

\section{Thermal Conductivity at Finite Frequencies}
For models involving particle flow, we recall that the heat current is defined as the energy current minus $\mu$ times the particle current. Therefore to generate the linear response equations,
we write the grand canonical ensemble Hamiltonian in the presence of a temperature variation as ($ \omega_c= \omega  + i 0^+$)
\beq
K=K_0+K_1 e^{- i \omega_c t}, \label{K}
\eeq
with adiabatic switching from the infinitely remote past $t=-\infty$ as usual, and $K_0= \sum_{\vec{r}} K(\vec{r}) = \sum_{\vec{r}} (H(\vec{r}) - \mu \ n(\vec{r}) )$. Here $H(\vec{r})$ is the energy density, and since we are mainly dealing with  lattice models, we sum over $r$. The operator  
\beq
K_1= \sum_{\vec{r}}  {\psi(\vec{r})} K(\vec{r}), \label{K1}
\eeq                        
where $\psi(\vec{r})$ is a small (pseudo)  gravitational field  with  some spatial variation such that its average is zero.  This expression follows  the conceptually important  work of Luttinger\cite{luttinger}. Alternate schemes such as the Kadanoff Martin method \cite{kadanoff} lead to the same results.
 Here  $\psi(r)$ is not quite a gravitational field since it couples to $K(r)$   rather than to $H(r)$, but serves the same purpose, it is a {\em mechanical field}  as opposed to the true spatially dependent temperature, which is a {\em thermodynamic field}. The latter is more subtle, and  ensuring  that such a varying field is established in a given system is in general a very hard thing to achieve rigorously- requiring as it does a detailed understanding of the equilibration processes that operate on a microscopic and inhomogeneous level. Luttinger's idea  reorganizes and subdivides  this complication into logically separate parts, and allows us to deal with a mechanical field with an arbitrary  spatial variation.  The connection with true temperature variation is the  next logical   step, where Luttinger showed that we may essentially regard $\nabla \psi(\ve{r})= \frac{\nabla T(\ve{r})}{T}$. Here we denote the local temperature by $T(\ve{r})$, and $\beta(\ve{r})= \frac{1}{k_B T(\ve{r})}$. 
 A formal discussion is provided by Luttinger\cite{luttinger}, who drew  an analogy with the Einstein relation. The latter   relates the response to the (mechanical)  electrical field to the observed  concentration gradient, which in turn is related to its conjugate namely  the   (thermodynamic) chemical potential gradient.  The resulting   frequency dependent linear response functions are  understood  as responses to this mechanical force. Relating these to  the thermodynamic response functions is the other part of the issue, and may be proved  only under certain  standard assumptions\cite{luttinger};  we are content to assume the same in this work. A simple qualitative argument  to help motivate the identification $\nabla \psi(\ve{r})= \frac{\nabla T(\ve{r})}{T}$, is that the temperature profile varies in such a way as to annul the variation of the added energy fluctuation, therefore if locally $\psi(\ve{r})$ increases so does $T(\ve{r})$ so as to maintain the local constancy of the action $\sum_{\ve{r}} \beta(\ve{r}) K(\ve{r})$.

Let  ${\cal L}$ be the number of sites, or equivalently  the volume of the system (by setting the lattice constant as unity). We next fourier transform  Eq(\ref{K1}) by introducing 
\barray
\psi(\ve{r})& =&  \sum_k \ha{\psi}(\ve{k}) \exp{( - i \ve{k}.\ve{r})}\\
K(\ve{r}) &= & \invol  \sum_k \ha{K}(\ve{k}) \exp( - i \ve{k} .\ve{r}),
\earray
so that 
\beq
K_1=  \sum_k \ha{K}(-\ve{k}) \ha{\psi}(\ve{k}). 
\eeq
The heat current operator  is Fourier decomposed as $\ve{J}^Q(\ve{r})= \invol \sum \exp{(- i \ve{k}. \ve{r})} \ve{\ha{J}^Q}(k)$.  
The   induced heat current in  response to the $k$th mode  with can be computed from linear response theory.  We will assume a ``temperature'' gradient  along the unit vector $\hat{x}$, so that $\Grad T(\ve{r}) = T \Grad \psi(\ve{r})= T \sum( - i k_x  \hat{x} )  \ha{\psi}(k_x) \exp{( - i k_x x )}$. Then
\barray
\kappa(\omega_c)& = & \lim_{k_x \rightarrow 0}\kappa(k_x,\omega_c) \;\mbox{with}\; \nonumber \\
 \kappa(k_x,\omega_c)& = &  \frac{ e^{ i \omega_c t}}{\cal{L}}  \left\{  \frac{\langle  \jq(k_x) \rangle }{ i T  k_x \ha{\psi}(k_x)}\right\}.  \label{kappa}
\earray
The expectation $\langle  \jq(k_x) \rangle$ is given by standard  linear response theory\cite{kubo,luttinger,fetter} and using Eq(\ref{kappa}) we find
\barray
 \kappa(k_x,\omega_c) & = & \frac{1}{ \hbar T k_x \cal{L}}  \int_{-\infty}^t \,  e^{i \omega_c (t- t')}\; dt' \langle[ \jq(k_x,t), \ha{K}(-k_x,t')]\rangle \label{kappa2} \\
& = &  \frac{-i }{\hbar \omega_c T k_x \cal{L}} \left( \langle[\jq(k_x),\ha{K}(-k_x)]\rangle + i \int_0^\infty \, e^{i \omega_c t'} \; dt' \langle[ \jq(k_x,t'), [\ha{K}(-k_x,0),K]]\rangle \right)\label{kappa3} 
\earray
We integrated by parts to get the second line from the first.
In the limit of almost uniform variation, $k_x\rightarrow 0$, we show that
\beq
 \langle[\jq(k_x),\ha{K}(-k_x)] \rangle    \sim  - k_x  \langle \Theta^{xx} \rangle, \label{observation}
\eeq
where  $\Theta^{xx}$ is the thermal  operator, to coin  a name.
In the uniform (i.e. $k_x \rightarrow 0$) operator $\jq$ is not a constant of motion,
and yet the average is of $O(k_x)$, this is so since the uniform term vanishes by 
noting  that the thermal average
$$
\langle [\jq,K] \rangle  \equiv  \frac{1}{\cal{Z}} \left[ Tr e^{- \beta K} \jq K - Tr e^{- \beta K}  K \jq \right] =0 \;\;\; \mbox{(Identity I)},
$$
with ${\cal{Z}}= Tr e^{- \beta K}$, 
the last identity following from the cyclicity of trace. Therefore we can write an equation for the {\em thermal operator} directly as
\beq
\Theta^{xx}= -\lim_{k_x\rightarrow 0}  \frac{d}{d k_x}[\jq(k_x),\ha{K}(-k_x)] \label{eqtheta}
\eeq
and thus it is straightforward if tedious  to compute it, given the explicit forms of the current and energy operators.

The heat current is obtained from the continuity equation for  heat density, written in momentum space as
\beq
{\lim_{k_x \rightarrow 0}} \frac{1}{k_x }[\ha{K}(-k_x),K]  = \hbar  \jq, \label{eqjq}
\eeq
where $\jq=\jq(k)/_{k\rightarrow 0}$,
so that the thermal conductivity at finite frequencies is
\beq
\kappa(\omega_c) = \frac{  i}{\hbar \omega_c T} \invol \left[ \langle \Theta^{xx} \rangle - i \int_0^\infty  e^{i \omega_c t'}\; dt' <[ \jq(t'), \jq(0)]> \right]. \label{kappa_final}
\eeq
It is worth making a comment on the difference between this calculation and that of $\sigma(\omega)$, the electrical conductivity at this point.
A  calculation of  $\sigma(\omega)$, as in  Appendix A, is in close parallel to this one for $\kappa(\omega)$, with the electrical current and  charge density fluctuation operators (defined in Appendix A) replacing the heat current and the energy density as in  $\jq(k_x) \rightarrow \jc(k_x)$ and with $\ha{K}(-k_x) \rightarrow \rho(-k_x)$.  In that case, the commutator $[\jc(k_x), \rho(-k_x)]$  explicitly vanishes as $k_x \rightarrow 0$, and hence the analog of the first term inside the round bracket  of Eq(\ref{kappa3}) automatically begins as  $O(k_x)$,  with a coefficient $\propto \langle \tau^{xx} \rangle $ ( defined in Eqn(\ref{tau})). In the present case, Eq(\ref{observation}) together with Eq(\ref{eqjq}) again leads to a cancellation of   $k_x$   between the numerator and the denominator, leading to a finite result in the uniform limit. Thus the thermal conductivity  calculation is somewhat disguised by the non-commutation of the uniform operators  i.e.  $[\jq,K] \neq 0$, which is luckily immaterial, since the {\em expectation of this object vanishes} due to the Identity I.

We perform a Lehmann representation \cite{fetter} to write Eq(\ref{kappa_final}) with $p_n =  \frac{1}{\cal Z} \exp( -\beta \epsilon_n )$ as
\beq
\kappa(\omega_c)= \frac{ i}{\hbar \omega_c T} \invol \left[  \langle \Theta^{xx} \rangle  - \hbar  \sum_{n,m} \frac{ p_n- p_m }{\epsilon_n - \epsilon_m + \hbar \omega_c} |\langle n|  \jq | m  \rangle|^2  \right]. \label{eq14}
\eeq
Using the  partial fractions identity (for any $\Delta$)
$$
\frac{1}{\hbar \omega_c (\hbar  \omega_c + \Delta) } = \frac{1}{\Delta} \left( \frac{1}{\hbar \omega_c}- \frac{1}{\hbar \omega_c + \Delta} \right),
$$
we obtain
\beq
\kappa(\omega_c)= \frac{  i}{\hbar \omega_c T} D_Q
+ \frac{  i \hbar}{ T \cal{L}}  \sum_{n,m} \frac{ p_n- p_m }{\epsilon_m - \epsilon_n} \frac{ |\langle n|  \jq | m  \rangle|^2}{ \epsilon_n - \epsilon_m + \hbar \omega_c}. \label{eqkubo1}
\eeq
where
\barray
 D_Q & =&  \invol \left[  \langle \Theta^{xx} \rangle  - \hbar \sum_{n,m} \frac{ p_n- p_m }{\epsilon_m - \epsilon_n} |\langle n|  \jq | m  \rangle|^2  \right],  \nonumber \\
&=&  \invol \left[  \langle \Theta^{xx} \rangle  - \hbar \int_0^\beta  d \tau \langle  \jq( - i \tau) \jq(0) \rangle \right]
\label{eqdq}
\earray 

Using  the standard identity $1/(\omega_c - E)= P \;1/(\omega- E) - i \pi \delta( \omega - E)$ for any $E$,
it is instructive to write the real part of the Lehmann representation of Eq(\ref{eqkubo1})  as
\barray
Re \; \kappa(\omega) &  = & \frac{\pi}{\hbar T}  \delta(\omega) \bar{D}_Q + Re \; \kappa_{reg}(\omega) \;\;\mbox{with} \nonumber \\
Re \; \kappa_{reg}(\omega)&=&  \frac{\pi}{   T \cal{L}} \left( \frac{1 - e^{-\beta \omega}}{\omega}\right) \sum_{\epsilon_n \neq \epsilon_m} p_n  |\langle n|  \jq | m  \rangle|^2 \delta( \epsilon_m- \epsilon_n -\hbar \omega), \label{eq16} \\
 \bar{D}_Q & = & \invol \left[  \langle \Theta^{xx} \rangle - \hbar \sum_{\epsilon_n \neq \epsilon_m} \frac{ p_n- p_m }{\epsilon_m - \epsilon_n} |\langle n|  \jq | m  \rangle|^2  \right]\label{eq17}.
\earray
Here $\kappa_{reg}$ is the regular part of thermal conductivity that excludes the delta function at $\omega=0$, and $\bar{D}_Q$ is
the weight of the delta function at  $\omega=0$. It is seen that the  weight of the zero frequency delta function  involves a different object than the one in Eq(\ref{eqdq}), there is a cancellation  of terms with equal energy between the two terms in Eq(\ref{eq14}). We elaborate on this distinction below. 

The sum  rule for the real part of the  thermal conductivity (an even function of $\omega$) follows  from Eq(\ref{eq16}) by integration, or equivalently from Eq(\ref{eq14}) and using the Kramers Kronig relation   to identify the coefficient of $i/\omega$ at high frequencies  in the complex conductivity as the integral of the real part:  
\beq
\int_{0}^\infty Re \; \kappa(\omega) d\omega = \frac{ \pi}{ 2 \hbar T \cal{L}} \langle \Theta^{xx} \rangle. \label{kappasumrule}
\eeq 
This interesting sum rule is one of the main formal results of this paper. The  extensive thermal operator $\Theta^{xx}$ is analogous to the stress tensor $\tau^{xx}$ in the sum rule Eq(\ref{fsumrule}) for the electrical conductivity $\sigma(\omega)$, and depends in its details upon the 
 underlying  model; we present it in the case of several problems of current interest in the following.

We may rewrite Eq(\ref{eqkubo1})  in a more compact form as
\beq
\kappa(\omega_c) = \frac{i }{ T \hbar \omega_c} D_Q  + \frac{1}{ T {\cal L}} \int_0^\infty dt e^{i \omega_c t } \int_0^\beta  d \tau \langle  \jq( -t - i \tau) \jq(0) \rangle. \label{kappaeq}
\eeq
The first term in Eq(\ref{kappaeq}) is $\propto D_Q$, and is in addition to the original Kubo formula\cite{kubo} for thermal conductivity, namely the second term. It represents non trivial a correction for non-dissipative situations, but vanishes in dissipative cases due to reasons that we elaborate below.

 Integrating the second term of Eq(\ref{kappaeq}) gives an estimate for 
$ \int_{0}^\infty Re \; \kappa(\omega) d\omega$ namely  $ \frac{ \pi}{ 2  T \cal{L}}  \int_0^\beta  d \tau \langle  \jq( - i \tau) \jq(0) \rangle $, which  has been written in literature earlier in its high temperature limit \cite{zotos}.  However it cannot be viewed as a sum rule, since the estimate is a correlation function of a pair of extensive currents,  rather than a direct expectation value of a single extensive operator as in  Eq(\ref{kappasumrule}). Further, in non-dissipative situations, this  estimate  misses the contribution from the $D_Q$ term and is thus incorrect. For dissipational systems, it is does reduce to  Eq(\ref{kappasumrule}), on using the vanishing of  $D_Q$  and Eq(\ref{eqdq}). Therefore it is clear that   in all cases, Eq(\ref{kappasumrule}) 
stands in complete parallel to the f-sum rule Eq(\ref{fsumrule}) for $\sigma(\omega)$ with the thermal operator $\Theta^{xx}$ relacing the stress tensor $\tau^{xx}$. 

We next make a few comments on the object $D_Q$, which
enters into the correct expression for thermal conductivity Eq(\ref{kappaeq}), and $\bar{D}_Q$ that enters in Eq(\ref{eq17}).  Note that the difference between $D_Q$ and $\bar{D}_Q$ lies in the nature of the double sum over the current matrix elements, $\bar{D}_Q$
explicitly excludes terms with equal energy, so that the difference is a sum over all degenerate manifolds (including  diagonal matrix elements $n=m$):
\beq
D_Q - \bar{D}_Q=  - \hbar \beta \sum_{\epsilon_n=\epsilon_m} p_n |\langle n | \jq | m \rangle|^2 \label{eq18}
\eeq
A detailed discussion of the charge stiffness in the similar context of charge transport is given in Ref\cite{giamarchi_bss}, where the distinction between zero and finite  temperatures is elaborated upon. Our discussion here is along similar lines, except that here we use the term charge stiffness to denote both $D_c$ and $\bar{D}_c$ whereas these are denoted by the terms  Meissner stiffness and charge stiffness in Ref\cite{giamarchi_bss}. In the case of the thermal conductivity $D_Q$ and $\bar{D}_Q$ are thus the thermal analogs of the Meissner stiffness and the charge stiffness of Ref\cite{giamarchi_bss}.

The  value of $D_Q$   is zero for most dissipational situations, i.e. for generic systems at finite temperatures. It is however non zero in general for systems that display macroscopic  many body coherence, i.e. either superfluidity or superconductivity.
For a superconductor, the analogous object, $D_c$ in the charge conductivity (see Eq(\ref{eqdc})) is in fact  related to the London penetration depth or the superfluid stiffness, and its non vanishing is the very hallmark of its ``super''-nature. It is alternately obtainable from the Byers-Yang relation $D_c \propto \lim_{\phi \rightarrow 0} \frac{d^2}{d \phi^2} F(\phi)$
where $F(\phi)$ is the free energy of the superconductor in the presence of a hole threaded by a flux $\phi$. While there is no obvious parallel to the flux for thermal conductivity, the formal expressions are very similar. 

There has been considerable discussion of the corresponding charge stiffness in recent literature\cite{integrable_charge}  for {\em integrable many body systems} in low dimensions, such as the Heisenberg or the Hubbard models in 1-d, where the existence of many conservation laws  pushes  them away  from  equilibrium with negligible restoring forces, leading to a ballistic  rather than diffusive behaviour, as first pointed out by Giamarchi\cite{integrable_charge}. An interesting numerical study of the statistics of the Kubo conductivity in 1-d has also given some insights around  the zero frequency limit\cite{huse}.  
Very recently the thermal conductivity has also been discussed in literature for integrable models\cite{heisenberg_1d}. In general terms, integrable models are akin to  the free particle system rather than to superconductors or superfluids, since there is no broken symmetry. In such cases, we expect $D_Q=0$ but it is  possible that $\bar{D}_Q \neq 0$, albeit possibly only for finite sized systems.
This is analogous to the charge stiffness story \cite{giamarchi_bss}, where $D_c$ is zero for normal metals whereas $\bar{D}_c$  may be non zero.   In this sense the zero temperature limit is singular;  $\bar{D}_c$  occurs in both conductivity and in the Byers-Yang- Kohn type formula $\bar{D}_c \propto \lim_{\phi \rightarrow 0} \frac{d^2}{d \phi^2} E_0(\phi)$, with $E_0(\phi)$ the ground state energy in the presence of a flux\cite{giamarchi_bss}, whereas at  non zero $T$ there is a  distinction as in the RHS of Eq(\ref{eq18}, \ref{eq189}). The  object $\bar{D}_Q$  involves  matrix elements of the current in states with {\em distinct energy}, and these are zero if the energy/heat current is  conserved, as in the case of the free electron system.  More non trivially, in the case of the exactly integrable Heisenberg model in 1-d,  the higher  conservation laws include the energy current  \cite{heisenberg_1d}. In such a case, since  the heat current is a constant of motion, it follows that  $\jq(t)=\jq(0)$,  leading to a simple calculation of $\kappa(\omega)$ using Eq(\ref{kappa_final}). 
 Here the second term in Eq(\ref{kappa_final})  vanishes on using $\jq(t)=\jq(0)$, giving
$$\kappa(\omega_c)= \frac{ i}{\hbar \omega_c T \cal{L}}    \langle \Theta^{xx} \rangle.  $$
Thus   $\kappa(\omega_c)$ may be computed for integrable models with a conserved energy current directly from a knowledge of  $\langle \Theta^{xx} \rangle$.

 We will show  in Sec IV  that the vanishing of $D_Q$ is  natural for generic systems, using  an argument based on the Kubo identity, and also pinpoint the possible technical reason as to why it is non zero for say a clean superconductor.  In subsequent sections we provide an evaluation of the thermal operator  $\Theta^{xx}$ given in  Eq(\ref{eqtheta}) for various  models of current interest.

\section{Thermopower}
We next study the linear response formulation for thermopower. If  we define the charge current and its Fourier decomposition as
 $\ve{J}(\ve{r})= \invol \sum \exp(- i \ve{k}.\ve{r}) \vec{\hat{J}}(\ve{k})$, then  in the presence of an electric field 
and a temperature gradient:
\beq
<\jc>=\sigma(\omega) E_x+ \gamma(\omega) (-\Grad T),
\eeq
where $\jc= \hat{J}_x(k_x \rightarrow 0)$.
 The thermopower is defined as
\beq
S(\omega)= \frac{\gamma(\omega)}{\sigma({\omega})}.
\eeq
We can obtain $\gamma(\omega)$ by linear response theory in  parallel to the thermal conductivity, and the answer is
\beq
\gamma(\omega_c)= \lim_{ k_x \rightarrow 0} \left(    \frac{ \langle \jc(k_x)\rangle e^{i \omega_c t}}{i T k_x \ha{\psi}(k_x) \cal{L}}   \right) .
\eeq
The expression for $\langle  \jc \rangle$ follows the same lines as that for the heat current and we find
\beq
\gamma(\omega_c)= \frac{i}{ \hbar \omega_c T \cal{L}} \left[ \langle \Phi^{xx} \rangle - i \int_0^\infty \; e^{i \omega_c t'} \langle [ \jc(t'),\jq(0)] \rangle \right]
\eeq
where we have introduced another important extensive operator, the ``thermoelectric operator'':
\barray
\Phi^{xx}&  \equiv &  - \lim_{k \rightarrow 0} \frac{d}{dk_x} [\jc(k_x),K(-k_x)]. \label{thermopowerphi}
\earray 
It is useful to perform its Lehmann representation, so that
\beq
\gamma(\omega_c)= \frac{ i}{\hbar \omega_c T \cal{L} }  \left[  <\Phi^{xx}> -  \hbar \sum_{n,m} \frac{ p_n- p_m }{\epsilon_n - \epsilon_m + \hbar \omega_c} \langle n|  \jc | m  \rangle \langle n|  \jq | m  \rangle  \right].
\eeq
We can proceed further as in the case of thermal conductivity by expanding  the product using partial fractions and find
\beq
\gamma(\omega_c)= \frac{ i}{\hbar \omega_c T } D_\gamma + \frac{i \hbar}{ T \cal{L}} \sum_{n,m} \left( \frac{ p_n- p_m }{\epsilon_m - \epsilon_n }\right)
\frac{\langle n|  \jc | m  \rangle \langle n|  \jq | m  \rangle  }{ \epsilon_n - \epsilon_m + \hbar \omega_c} .
\eeq
where
\beq
D_\gamma=  \invol \left[  <\Phi^{xx}> - \hbar \sum_{n,m} \frac{ p_n- p_m }{\epsilon_m - \epsilon_n } \langle n|  \jc | m  \rangle \langle n|  \jq | m  \rangle \right]. 
\eeq
Since the two currents involved are not identical, it is not possible to take the real part of this expression readily, and it is not profitable to seek a sum rule. It is however, interesting to note that by combining various formulae, the thermopower has a high frequency expansion
of the type
\beq
S(\omega)= S^* + O(\frac{1}{\omega})\;\; \mbox{where}\;\; S^*= \frac{\langle \Phi^{xx}\rangle}{T \langle \tau^{xx} \rangle}. \label{sstar}
\eeq
and so a knowledge of $\Phi^{xx}$ is useful in determining  the asymptotic behaviour.
 We can rewrite this equations as
\beq
\gamma(\omega_c) = \frac{i }{ \hbar \omega_c T} D_\gamma  + \frac{1}{ T {\cal L}} \int_0^\infty dt e^{i \omega_c t } \int_0^\beta  d \tau \langle  \jc( -t - i \tau) \jq(0) \rangle, \label{eqgamma}
\eeq 
We again note that $D_\gamma=0$ for most dissipational cases.

\section{Kubo type formula for general non dissipative systems}
In  an earlier section,  we have derived  the linear response theory expressions for the various conductivities Eqs(\ref{kappa}), and 
argued that the usual formulae quoted in literature e.g. in Ref\cite{mahan} following Kubo\cite{kubo} and others\cite{luttinger} have extra terms that 
arise for non dissipative systems, such as superconductors and superfluids. The reader might wonder as to the origin of these extra terms, since the common belief is that the Kubo formulas are complete and exact. In this section we obtain these terms from a careful analysis of the Kubo arguments. We deal with the case of thermal conductivity below, but the arguments apply equally to the electrical conductivity as the reader can easily see. In  the case of the electrical conductivity a simpler  argument gives the final result using the transverse gauge, i.e. by realizing  the  electrical field as the time derivative of a vector potential rather than as the gradient of a scalar potential\cite{martin,ss}, although the present method is readily generalized to that case as well, as shown in the Appendix. 

We begin with Eq(\ref{kappa2}) rearranged in the form preferred by Kubo ;
\beq
\kappa(k_x,\omega_c )= \frac{1}{k_x T \cal{L}} \int_0^\infty dt\; e^{i \omega_c t} Tr\left( [\rho_0,\hat{K}(-k_x,- t)] \jq(k_x) \right),
\eeq
with the density matrix  $\rho_0= \frac{e^{-\beta K}}{{\cal Z}}$.
 From this point onwards one uses  three important steps 
\begin{enumerate}
\item The so called ``Kubo identity'' \cite{kubo} which is written as  ${\cal{K}}[K(-k_x,-t)] = 0$ where
\beq {\cal{K}}[A] \equiv [\rho_0, A] - \int_0^\beta \rho_0 [ A(-i \tau) ,K] d\tau ,
\eeq   
which is easily derived by inserting a complete set of states.
\item The conservation law 
\beq \frac{1}{\hbar k_x} [\hat{K}(-k_x,- t),K]= \jq(-k_x,-t) \label{conservation}.  \eeq
\item The homogenous or uniform limit    $k_x \rightarrow 0$.  
\end{enumerate}
These three steps together give the usual stated result  Eq(\ref{kappa_final}) without the first term involving $D_Q$.  These  steps lead one to an expression involving  the matrix elements of $\hat{K(-k_x,-t)}/k_x$ and  it is tempting   to replace this   with those of $\jq(0,-t)$. It is precisely  here that one needs to be careful.
As long as $k_x$ is finite it is clear that $\hat{K}(-k_x,-t)$ as well as $\jq(-k_x, -t)$ have matrix elements between eigenstates of the Hamiltonian $K$ that are at {\em different energies}, i.e. inhomogeneous excitations cost energy. In the limit of $k_x \rightarrow 0$ this need not be so, the current operator {\em can have} matrix elements between states of the same energy, whereas by construction the commutator $[ \hat{K}(-k_x,-t),K]$ filters out states with the same energy.  Thus the operator identity Eq(\ref{conservation}) is     
potentially  in trouble precisely where we need it.  What is
easier to justify  is to use a time derivative to filter out equal energy states so that a good  alternative identity to Kubo's identity is to use
\beq
 \frac{d  {\cal{K}}[K(-k_x, -t)]}{ d t} = 0. \label{new_identity_1}
\eeq
This  is the Kubo identity again, but with the operator $(-i) [\hat{K}(-k_x,-t),K]$ in place of $\hat{K}(-k_x,-t)$, this extra commutator with $K$ is transferred to $\jq(-k_x,-t)$ on using the conservation law, and improves matters. In this form we can take the next two steps safely since $[\jq(0,-t),K]$ does not have any matrix elements between equal energy states.  We can now integrate the 
Eq(\ref{new_identity_1}) between finite times and get 
\beq
{\cal{K}}[K(-k_x, -t)]= {\cal{K}}[K(-k_x, 0)] \label{new_identity_2},
\eeq
This equation tells us that the errors in replacing the matrix elements of $\hat{K(-k_x,-t)}/k_x$ with those of $\jq(-t)$ are {\em time independent}, i.e. relating solely to the matrix elements of current in the  manifold of zero energy difference states. Such a manifold of states is statistically significant only in a superconductor, and hence we expect it to be relevant in that context.

Proceeding as before to the uniform limit, we get 
\barray
\kappa(\omega_c) & = & \frac{i}{ \hbar \omega_c T \cal{L}} \left\{ \langle \Theta^{xx} \rangle - \hbar \int_0^\beta d\tau \langle \jq(- i \tau) \jq(0) \rangle \right\} \nonumber \\
 && + \frac{1}{  T \cal{L}} \int_0^\infty e^{ i \omega_c t} dt\;  \int_0^\beta d\tau \langle \jq(-t - i \tau) \jq(0) \rangle.
\earray
The term in curly brackets  is recognized as ${\cal{L}}D_Q$ as in Eq(\ref{kappaeq}), and hence the two derivations lead to exactly the same answer.  

We have seen that $D_Q$ owes its possible non vanishing to the matrix elements of the current operator in the  manifold of zero energy difference states, such as exist in a superconductor. For most dissipative systems, such a manifold is statistically insignificant,  and   it should be noted that in such  instances, $D_Q$ is actually zero. 

The only instances where $D_Q$ and analogously $D_c$ of Eq(\ref{eqdq}, \ref{eqdc})  are non zero have to do with superfluid or superconducting states.  We remind the reader of  the change in terminology from  Ref\cite{giamarchi_bss} as remarked on earlier below Eq(\ref{eq18}),
 $D_c$ represents the Meissner stiffness of  Ref\cite{giamarchi_bss}.  In these cases, $D_c$ for example is proportional to the superconducting density of electrons $\rho_s$ as in the London theory. 
Traditionally  $\rho_s$ is obtained by taking the current current correlator and taking the limit $\ve{k} \rightarrow 0$ for the {\em transverse } part, i.e. $\ve{J} .\ve{k}=0$, after having taken the static limit at the outset- this is the classic Meissner effect calculation\cite{meissnereffect}.   However, it is also natural to see the superconducting fraction in the 
 conductivity where we first set $\ve{k} \rightarrow 0$ and {\em then let } $\omega \rightarrow 0$. 
This is the procedure followed in the well known Ferrell-Glover-Tinkham sum rule determination of the penetration depth in superconductors from the infrared conductivity\cite{ferrell}.  In this sequence of taking the  uniform limit, it  should  not matter  if we have worked with the longitudinal  conductivity rather than the transverse one, and our derivation here is the longitudinal counterpart of the transverse derivation first obtained   for the Hubbard type models  in Ref(\cite{ss}). A nice discussion of the various limits is given in Ref(\cite{scalapino}). 
A superconductor would also have a non zero $D_Q$ if impurities are neglected, however impurities {\em do scatter} quasi particles and hence one expects disorder to make this object tend to vanish, unlike the variable $D_c$ which is  less sensitive to disorder by virtue of the Anderson theorem. There is also the similarity of the $D_Q$ term with that of second sound that one expects in a superconductor, namely an undamped energy wave; $D_Q$ then appears as the residue at a second sound pole in the uniform limit of a suitable propagator.  

This derivation also indirectly  shows that in the normal dissipative cases, when $D_Q=0$, the sum  rule Eq(\ref{kappasumrule}) is obtained starting from the conventional Kubo formula, i.e. Eq(\ref{kappaeq}) without the first term, by integrating the current current correlator over frequency. This is so  since such an integral gives the second part of the expression for $D_Q$ which must equal the first for  it to vanish!   

It is worth mentioning that another independent argument for the vanishing of $D_Q$ and likewise $D_c$, or in another context, the spin wave stiffness of a fully saturated ferromagnet\cite{herring}, uses the notion of a position operator. If a position operator, or a similar one for energy transport such as the boost operator ( $\int \; dr \ve{r} H(r)$)  is allowed in the list of allowed operators, then it is easy to show that $D_c$ type operators are zero, since one can rewrite the current matrix elements in terms of the matrix elements of the position operator, and then the second part of  Eq(\ref{eqdc}) becomes the commutator of the current with the position operator and hence cancels exactly the $\tau^{xx}$ term. Such a procedure is invalid in general, as explained carefully by Herring and Thouless\cite{herring}, the position operator is illegal in the Hilbert space where periodic boundary conditions are used in a continuum field theory.   For a lattice field theory of the sort we are considering, we can always force the  introduction of a lattice position operator ${\cal X} \equiv \sum_{\ve{r}} x_j n_{\ve{r}_j}$, but now a careful calculation shows that the cancellation is incomplete, while the bulk terms do cancel, the boundary terms do not.

From this line of thought we conclude that $D_c$ and its relatives would vanish, if the nature of the quantum states is such that the  error made in introducing the position operator is negligible, this again is plausible in a highly dissipative system, where phase coherence is lost over some microscopic length scale that is shorter than the system length.  For a superconductor such is certainly not the case, and nor for that matter, in   a ferromagnet Ref(\cite{herring}).

\section{High frequency response: Lorentz ratio and Thermoelectric Figure of Merit}

Of great interest are the two quantities, the Lorentz ratio
\beq
{\bf L}(\omega) = \frac{\kappa(\omega)}{ T \sigma(\omega)}
\eeq
and the thermoelectric figure of merit $Z$ times $T$ (a dimensionless number)
\barray
{\bf Z}(\omega) T & = & \frac{S^2(\omega) \sigma(\omega) T }{ \kappa(\omega) }\\
&=& \frac{\gamma^2(\omega) T }{  \sigma(\omega) \kappa(\omega)}. \label{powerfactor}
\earray

We can readily evaluate these in the high frequency limit as
\barray
{\bf L}^*& = & \frac{ \langle \Theta^{xx} \rangle }{T^2 \langle \tau^{xx} \rangle } \\
{\bf Z}^* T &=&  \frac{ \langle \Phi^{xx} \rangle^2   }{ \langle \Theta^{xx} \rangle  \langle \tau^{xx} \rangle }.
\earray

Let us also note that the thermopower at high frequencies is expressible in terms of the operators we displayed earlier:
\beq
S^*= \frac{\langle \Phi^{xx}\rangle}{T \langle \tau^{xx} \rangle}.
\eeq
These variables are computable much more readily than their dc counterparts, and since these do capture the effect of interactions, they are of considerable interest. 

\section{Lattice Thermal Conductivity }

We consider here the sum rule as applied to the case of the lattice vibrations.
This is a field of great current activity \cite{thermal_lattice,onuttom}, where most studies are in low dimensions and treat classical anharmonic and disordered lattices with a view to study the conditions for existence of a Fourier law, and also for the emergence of a finite conductivity. 
 We consider a general anharmonic disordered lattice
with a Hamiltonian
\barray
H & = & \sum_j H_j \nonumber \\
H_j&=&   \left[ \frac{\ve{p}_j^2}{2 m_j} + U_j \right];  \hspace{.3in}
U_j =  \frac{1}{2} \sum_{i \neq j} V_{j,i},
\earray
where $V_{i,j}=V_{j,i}=V(\ve{u}_i-\ve{u}_j)$ is a symmetric two body potential that is an arbitrary   function of the displacement variables $\ve{u}_i$ including possible anharmonicity.
Since no particle flow is involved, it suffices to consider the canonical ensemble Hamiltonian.
 We will denote the equilibrium lattice points by $\ve{R}_j\equiv( X_j,Y_j,..)$ and the antisymmetric force vector $\ve{F}_{i,j}= - \frac{\ve{\partial}}{\partial \ve{u}_i}V_{i,j}$. In the following, we will need the derivatives of the antisymmetric force vector which will be denoted by
$F^{x (n)}_{i,j}= \frac{\partial^n}{\partial (u^x_{i})^n} F^x_{i,j}$, and have the obvious symmetry property $F^{x (n)}_{j,i}=(-1)^{n-1} F^{x (n)}_{i,j}$. The energy current can be obtained from computing the time evolution of the energy density, and combining it with the conservation law thereby giving 
 \beq
J^E_x= \lim_{k\rightarrow 0} \frac{1}{\hbar k_x} [ H, H(k)]
\eeq
where $H(k)= \sum_j e^{ i \ve{k}.\ve{R}_j} H_j$. This form of the current operator is popular in literature, e.g. Ref(\cite{grossberger})  and differs from the one given by Hardy \cite{hardy} in some details, which are not expected to be very significant. To aid the passage to a classical limit we display the dependence on  Planck's constant in this section, setting it to unity elsewhere. The current at finite wave vectors can be found from the above as
\beq
 J^E_x(\ve{k})=\frac{1}{4 } \sum_{i,j} \frac{(X_i-X_j)}{m_i} e^{ i k_x X_i} \left\{ p_{x,i}, F^x_{i,j} \right\}
\eeq
where  we have chosen $\ve{k}$ along the $x$ axis. Using the various definitions, we find the thermal operator
\barray
\frac{1}{\hbar} \Theta^{xx}& =& \sum_i \frac{1}{4 m_i}\left[ \sum_{j\neq i}F^x_{i,j} (X_i-X_j) \right]^2 \nonumber \\
&&- \frac{1}{8} \sum_{i,j} \frac{(X_i-X_j)^2}{m_i m_j} \left[  4 F^{x(1)}_{i,j} \;p_{x,i}\; p_{x,j} + 2 i \hbar\; F^{x (2)}_{i,j}\; (p_{x,i}-p_{x,j}) + \hbar^2\; F^{x(3)}_{i,j} \right].
\earray
This together with Eq(\ref{kappasumrule}) then gives us the sum rule for thermal conductivity. Note that the  classical limit
exists and is readily found by dropping terms involving $\hbar$  in this expression.  In the classical limit of Kubo's formula Eq(\ref{kappaeq}), the inner integral over $\tau$ collapses to give 
\beq
\kappa_{classical}(\omega_c)= \frac{1}{k_B T^2 \cal{L}} \int_0^\infty \; e^{ i \omega_c t} \; dt\; \langle \jq(-t) \jq(0) \rangle_{classical}  \label{classical_kubo},
\eeq
and it is clear that we can express the sum rule in terms of  the expectation of $\langle \jq \jq \rangle $. However our formula Eqn(\ref{kappasumrule}) provides a slight advantage in this case too: our result implies that
 \beq
\frac{1}{\hbar} \langle \Theta^{xx} \rangle_{classical} = \frac{1}{k_B T} \langle \jq \jq \rangle_{classical},
\eeq  
a result that is non trivial  since the LHS involves a single extensive operator whereas the RHS involves the square of an extensive operator, and moreover the only way to prove it seems to be to go back to the inhomogeneous (i.e. k dependent) energy density and perform the steps stated  above!

 As a simple illustration consider the  case  1-d Harmonic lattice, $H=\sum_i p_i^2/(2 m) + k_s/2 \sum_i( u_i-u_{i+1})^2$, with $k_s \equiv  m \omega_0^2$.
In this case, we expect heat transport to be ballistic rather than diffusive, and hence the static thermal conductivity to diverge.
Unlike the case of free electrons  (discussed elsewhere in this paper), the current operator is not completely diagonal in the normal mode (phonon) operators, in contrast to the Peierls form of the energy current\cite{peierls}, namely $ J^E= \sum \hbar \omega_k v_k n_k$.     The integrated conductivity is non trivial, it gets its weight at finite frequencies from phonon pair creation and destruction processes. The force is linear in the displacements and hence we work out the thermal operator:
\beq
\Theta^{xx}= (\hbar \omega_0^2 a_0^2) \left[ \frac{1 }{  m} \sum_i p_i p_{i+1} + \frac{ k_s}{ 4} \sum_i(u_{i-1}-u_{i+1})^2 \right].
\eeq
The expectation of this object can be computed  at finite temperatures on using the phonon harmonic oscillator representation as 
\beq
\langle \Theta^{xx} \rangle = {\cal{L}} (\hbar^2 \omega_0^3 a_0^2) \int_0^\pi \frac{ dk}{\pi} \left[  (\frac{1}{2}+\frac{1}{ e^{\beta \omega_k}-1}) \left\{ \frac{\omega_k}{\omega_0} \cos(k) + \frac{\omega_0}{\omega_k} \sin^2(k) \right \} \right],
\eeq
where $\omega_k= 2 \omega_0  \sin(k a_0/2)$ is the acoustic phonon energy.
At zero temperature, i.e. in the ground state  it can be  shown to vanish exactly; this precise cancellation  is a characteristic  feature of the thermal operators that is in common with most other quantum systems, and we discuss its connection with the vanishing of the specific heat in another section (see discussion near Eq(\ref{specific_heat})). Similar expressions can be worked out in any dimension $d$. At low temperatures, it is easy to see that $\Theta^{xx} \propto T^{d+1}$ in $ d$ dimensions, so that the sum rule Eq(\ref{kappasumrule}) is $\propto T^d$,  i.e. similar to  the lattice specific heat. At high temperatures, since this model has no scattering of phonons, the sum rule increases linearly with $T$ indefinitely.  Through Eq(\ref{kappasumrule}) this implies a $T$ independent behaviour at sufficiently high $T$. The computation for the non linear lattice  would contain the effects of umklapp as well as normal scattering, and should be  numerically feasible as well as interesting.  

\section{Intermediate coupling models:\\
 Finite U Hubbard model,  Homogeneous Electron Gas and Periodic Anderson Lattice   }

In this section we will consider the popular intermediate coupling models, namely the Hubbard model and the electron gas as well as the Periodic Anderson Lattice. We derive for these models the Thermal operator  $\Theta^{xx}$and the Thermoelectric operator  $\Phi^{xx}$. We further demonstrate the formalism in the limit of zero interaction, where the Boltzmann Drude results are reproduced for the  specific transport 
variables mentioned in the earlier sections, namely the Lorentz number and the thermopower. 
We begin by writing  the Hubbard model Hamiltonian in the grand canonical ensemble
\barray
K & = & \sum_{\ve{r}} K(\ve{r}) \\
K(\ve{r}) &=&  - \sum_{\sigma,\ve{\eta}} t(\ve{\eta}) c^\dagger_{ \ve{r}+ \frac{1}{2}\ven, \sigma} c_{ \ve{r} - \frac{1}{2} \ven, \sigma}
- \mu \sum_\sigma n_{\ve{r} \sigma} + U n_{\ve{r} \downarrow} n_{\ve{r} \uparrow}.
\earray
Here  $\ven$ represents all the  neighbors of the site $\ve{r}$, $t(\ven)$ is the hopping matrix element, and we have adopted the ``midpoint rule'' for defining densities of all non local objects such as hopping so as to get convenient expressions for the fourier transforms of the grand canonical ensemble Hamiltonian $K= H - \mu \hat{N}$
\barray
\hat{K}(\ve{k}) & = & \sum_{\ve{r}} e^{ i \ve{k}.\ve{r}} K(\ve{r})\\
&=&\sum_{\ve{p},\sigma} (\varepsilon_{\ve{p}} - \mu) \; c^\dagger_{\ve{p}+ \frac{1}{2} \ve{k}, \sigma} c_{\ve{p} - \frac{1}{2} \ve{k}, \sigma} + \sum_{\ve{r}} e^{ i \ve{k}.\ve{r}}   U n_{\ve{r} \downarrow} n_{\ve{r} \uparrow} , \label{khat}
\earray
with $\varepsilon_{\ve{p}} = - \sum_{\ven} t(\ven) \exp{( - i \ven . \ve{p})}$ the kinetic energy and 
$n_{\ve{r} }= \sum_\sigma n_{\ve{r},\sigma }$. A more general model, with arbitrary two body interaction between particles, including the homogeneous electron gas  (HEG) model is represented by
\beq
\hat{K}(\ve{k}) = \sum_{\ve{p},\sigma} (\varepsilon_{\ve{p}} - \mu) \; c^\dagger_{\ve{p}+ \frac{1}{2} \ve{k}, \sigma} c_{\ve{p} - \frac{1}{2} \ve{k}, \sigma} + \frac{1}{2 \cal{ L}} \sum_{\ve{p},\ve{q},\ve{l},\sigma,\sigma'} U(\ve{q}) c^\dagger_{\ve{p}+\ve{q}+\frac{1}{2}\ve{k} ,\sigma} c_{\ve{p}- \frac{1}{2} \ve{k} ,\sigma} c^\dagger_{\ve{l}- \ve{q} ,\sigma'} c_{\ve{l} ,\sigma'}
\eeq
By specializing to $U(\ve{q})=U$ and restricting the momenta to the first Brillouin Zone we recover the Hubbard model, whereas
the electron gas is obtained by letting $U(\ve{q})=4 \pi e^2/|\ve{q}|^2$ and letting ${\cal{L}}\rightarrow \Omega$ the box volume.
In all cases, the electronic charge density is $\rho(\ve{r})= q_e n_{\ve{r}}$ where $q_e$ is the charge of the electron ($q_e= - |e|$). 
The charge current density is given by $\ve{J}(\ve{r})= i \frac{q_e}{\hbar} \sum_{\ven, \sigma}\; \ven t(\ven) \; c^\dagger_{\ve{r}+ \frac{1}{2} \ve{\eta}, \sigma} c_{ \ve{r}- \frac{1}{2} \ve{\eta}, \sigma}$, and hence its Fourier transform follows as
\beq
\ve{\hat{J}}(\ve{k}) = q_e \sum_{\ve{p}} \ve{v}_{\ve{p}} \; c^\dagger_{\ve{p}+ \frac{1}{2} \ve{k}, \sigma} c_{\ve{p} - \frac{1}{2} \ve{k}, \sigma},
\eeq
where the velocity vector $\ve{v}_{\ve{p}}= \frac{\partial}{\partial \hbar \ve{p}} \;\ep =\frac{ i}{\hbar} \sum_{\ven} t(\ven) \ven \exp{(- i \ven .\ve{p})}$. We set the lattice constant $a_0=1$ throughout this paper.

\subsection{The Thermoelectric operator $\Phi^{xx}$}

Let us now work out the thermo electric term $\Phi^{xx}$ defined in Eq(\ref{thermopowerphi}).  The commutator can be taken and on further carrying out the limiting procedure we find with $\bar{\sigma} = - \sigma$
\barray
\Phi^{xx}& = & -\frac{q_e}{2 \hbar} \sum_{\ven,\vec{\eta'},\ve{r},\sigma} ( \eta_x+\eta'_x)^2 t(\ven) t(\venp) c^\dagger_{\ve{r}+\ven+\venp,\sigma} c_{\ve{r},\sigma} - q_e \frac{\mu}{\hbar}  \sum_{\ven,\sigma} \eta_x^2 t(\ven) c^\dagger_{\ve{r}+\ven,\sigma} c_{\ve{r},\sigma}+ \nonumber \\
&& \frac{ q_e U}{4 \hbar} \sum_{\ve{r},\ven,\sigma} t(\ven) (\eta_x)^2 ( n_{\ve{r}, \bar{\sigma}} + n_{\ve{r}+\ven, \bar{\sigma}}) ( c^\dagger_{\ve{r}+\ven,\sigma} c_{\ve{r},\sigma}+ c^\dagger_{\ve{r},\sigma} c_{\ve{r}+\ven,\sigma} ). \label{phihubbard1}
\earray
This object can be expressed completely in Fourier space as
\beq
\Phi^{xx} = q_e \sum_{\ve{p},\sigma} \frac{\partial }{\partial p_x} \left\{   v^x_p( \ep- \mu)  \right\} c^\dagger_{\ve{p},\sigma } c_{\ve{p},\sigma} + \frac{ q_e U}{2 \cal{L} } \sum_{\ve{l},\ve{p}, \ve{q},\sigma }  \frac{\partial^2 }{\partial l_x^2} \left\{ \varepsilon_{\ve{l}}+\varepsilon_{\ve{l}+\ve{q}} \right\} c^\dagger_{\ve{l}+\ve{q}, \sigma} c_{\ve{l}, \sigma} c^\dagger_{\ve{p}- \ve{q} ,\bar{\sigma}} c_{\ve{p}, \bar{\sigma} } \label{phihubbard}.
\eeq
The more general HEG model yields the result: 
\barray
\Phi^{xx}& = & q_e \sum_{\ve{p},\sigma} \frac{\partial }{\partial p_x} \left\{   v^x_p( \ep- \mu)  \right\} c^\dagger_{\ve{p},\sigma } c_{\ve{p},\sigma}  \nonumber \\
& &+ \frac{ q_e }{2 \cal{L} } \sum_{\ve{l},\ve{p}, \ve{q},\sigma,\sigma' } \left[ U(\ve{q}) \frac{\partial^2 }{\partial l_x^2} \left( \varepsilon_{\ve{l}}+\varepsilon_{\ve{l}+\ve{q}} \right) + \frac{\partial U(\ve{q}) }{\partial q_x} \left(  v^x_{\ve{l}+\ve{q}}- v^x_{\ve{l}} \right) \right]c^\dagger_{\ve{l}+\ve{q}, \sigma} c_{\ve{l}, \sigma} c^\dagger_{\ve{p}- \ve{q} ,{\sigma'}} c_{\ve{p}, {\sigma'} } .
\earray

\subsection{ Heat Current and The Thermal Operator $\Theta^{xx}$}
We first derive the expression for the heat current: this follows from Eq(\ref{eqjq}). Using Eq(\ref{khat}) and the fact that the $O(U^2)$ term vanishes (  both terms are functions of $n_{\ve{r},\sigma}'s$ only), we find
\barray
J^{Q}_x & = & \sum_{\ve{p},\sigma}  v^x_{\ve{p}} \left( \varepsilon_{\ve{p}} - \mu\right) \; c^\dagger_{\ve{p},\sigma } c_{\ve{p},\sigma} +
\frac{ i U}{2 \hbar } \sum_{\ven,\sigma} t(\ven) \eta_x  c^\dagger_{\ve{r}+ \ve{\eta}, \sigma } c_{ \ve{r}, \sigma } \left\{  n_{\ve{r}, \bar{\sigma}} + n_{\ve{r}+\ven, \bar{\sigma}}   \right\}\\
&=& \sum_{\ve{p},\sigma} v^x_{\ve{p}} \left(   \varepsilon_{\ve{p}} - \mu\right)\; c^\dagger_{\ve{p},\sigma } c_{\ve{p},\sigma} + \frac{U}{2 \cal{L}} \sum_{\ve{l},\ve{p},\ve{q},\sigma ,\sigma' }\left\{ v^x_{\ve{l}}+ v^x_{\ve{l}+\ve{q}} \right\} c^\dagger_{\ve{l}+\ve{q}, \sigma} c_{\ve{l}, \sigma} c^\dagger_{\ve{p}- \ve{q} ,{\sigma'}} c_{\ve{p}, {\sigma'} }
\earray
We also need the current operator at a finite wave vector chosen along the $x$ axis as $\ve{k}\equiv  \hat{x}\; k_x$, this may be written
in the case of the more general HEG model  as
\barray
J^{Q}_x(\ve{k}) & = & \sum_{\ve{p},\sigma} v^x_{\ve{p}} \left( \varepsilon_{\ve{p}} - \mu\right)  \; c^\dagger_{\ve{p}+\frac{1}{2} \ve{k},\sigma } c_{\ve{p}-\frac{1}{2} \ve{k},\sigma} + \frac{1}{2 \cal{L}} \sum_{\ve{l},\ve{p},\ve{q},\sigma ,\sigma' } U(\ve{q}) \left\{ v^x_{\ve{l}}+ v^x_{\ve{l}+\ve{q}} \right\} c^\dagger_{\ve{l}+\ve{q}+\frac{1}{2} \ve{k}, \sigma} c_{\ve{l} -\frac{1}{2} \ve{k}, \sigma} c^\dagger_{\ve{p}- \ve{q} , {\sigma'}} c_{\ve{p},  {\sigma'} }
\earray

We evaluate the thermal operator by a direct calculation:
\barray
&& \Theta^{xx} =  \sum_{p,\sigma} \frac{\partial}{\partial p_x} \left\{ v^x_{\ve{p}} (\varepsilon_{\ve{p}} - \mu)^2 \right\} \; c^\dagger_{\ve{p},\sigma} c_{\ve{p},\sigma}  \nonumber \\
&&+ \frac{1}{2 \cal{L}} \sum_{\ve{p},\ve{q},\sigma} U(\ve{q})
\left[  \frac{\partial  }{\partial p_x} \left\{ \xi_{\ve{p}}+ \xi_{\ve{p}+\ve{q}}       \right\} + \frac{1}{2} ( v^x_{\ve{p}}+ v^x_{\ve{p}+\ve{q}})^2 \right]
 c^\dagger_{\ve{p}+\ve{q},\sigma} c_{\ve{p},\sigma} \rho_{-\ve{q}}  \nonumber  \\
&& + \frac{1}{ 4 \cal{L}} \sum_{\ve{p},\ve{q},\ve{l},\sigma,\sigma'} 
 [ U(\ve{q}) \left\{ (\varepsilon_{\ve{l}-\ve{q}} - \varepsilon_{\ve{l}}) \frac{ \partial }{ \partial p_x} (v^x_{\ve{p}}- v^x_{\ve{p}+\ve{q}}) +
(v^x_{\ve{p}}+ v^x_{\ve{p}+\ve{q}}) (v^x_{\ve{l}} + v^x_{\ve{l}- \ve{q}}) \right\}  \nonumber \\
& & - 2 \frac{\partial U(\ve{q})}{ \partial q_x} (v^x_{\ve{p}} + v^x_{\ve{p}+\ve{q}}) (\varepsilon_{\ve{l}-\ve{q}} - \varepsilon_{\ve{l}}) ] \; c^\dagger_{\ve{p}+\ve{q},\sigma} c_{\ve{p},\sigma}  c^\dagger_{\ve{l}- \ve{q},\sigma'} c_{\ve{l},\sigma'}  \nonumber \\
&&+   \frac{1}{4 {\cal{L}}^2} \sum_{\ve{p},\ve{q},\ve{q'},\sigma} U(\ve{q}) [ U(\ve{q'})  \frac{\partial  }{\partial p_x} 
\left( v^x_{\ve{p}+\ve{q}+\ve{q'}}+  v^x_{\ve{p}+\ve{q}} + v^x_{\ve{p}+\ve{q'}} + v^x_{\ve{p}}\right) 
 \nonumber \\
&& +   \frac{ \partial U(-\ve{q'})}{\partial q'_x} 
\left( v^x_{\ve{p}+\ve{q}+\ve{q'}}+  v^x_{\ve{p}+\ve{q'}} - v^x_{\ve{p}+\ve{q}} - v^x_{\ve{p}}\right)
] c^\dagger_{\ve{p}+\ve{q}+\ve{q'},\sigma} c_{\ve{p},\sigma}  \rho_{-\ve{q}} \rho_{-\ve{q'}}.  
\earray
Here we denote $\xi_{\ve{p}}= v^x_{\ve{p}} ( \varepsilon_{\ve{p}} - \mu)$ and the density fluctuation  $\rho_{\ve{q}}= \sum_{\ve{p}} c^\dagger_{\ve{p}+\ve{q},\sigma} c_{\ve{p},\sigma}   $. This expression simplifies somewhat in the case of the Hubbard model in real space with $U(\ve{q}) \rightarrow U$, where it may be written as: 
\barray
&& \Theta^{xx} =  \sum_{p,\sigma} \frac{\partial}{\partial p_x} \left\{ v^x_{\ve{p}} (\varepsilon_{\ve{p}} - \mu)^2 \right\} \; c^\dagger_{\ve{p},\sigma} c_{\ve{p},\sigma} +\frac{U^2}{4 \hbar} \sum_{\eta,\sigma} t(\ve{\eta}) \eta_x^2 ( n_{\ve{r},\bar{\sigma}}+ n_{\ve{r}+\ven ,\bar{\sigma}})^2 c^\dagger_{\ve{r}+\ven ,{\sigma}} c_{\ve{r},\sigma}  \nonumber \\
&&
-  \mu \frac{U}{\hbar}  \sum_{\ven,\sigma} t(\ven) \eta_x^2 ( n_{\ve{r},\bar{\sigma}}+ n_{\ve{r}+\ven ,\bar{\sigma}}) c^\dagger_{\ve{r}+\ven ,{\sigma}} c_{\ve{r},\sigma}  \nonumber \\
&&- \frac{U}{8 \hbar} \sum_{\ven,\ven',\sigma} t(\ven) t(\ven') (\eta_x+\eta'_x)^2 \left\{ 3 n_{\ve{r},\bar{\sigma}}+ n_{\ve{r}+\ven ,\bar{\sigma}}+ n_{\ve{r}+\ven',\bar{\sigma}}+  3 n_{\ve{r}+\ven+\ven' ,\bar{\sigma}} \right\} c^\dagger_{\ve{r}+\ven+\ven' ,{\sigma}} c_{\ve{r},\sigma}  \nonumber \\
&& + \frac{U}{4 \hbar} \sum_{\ven,\ven',\sigma} t(\ven) t(\ven') (\eta_x+\eta'_x) \eta'_x  
c^\dagger_{\ve{r}+\ven ,{\sigma}} c_{\ve{r},\sigma} \left\{ c^\dagger_{\ve{r}+\ven ,\bar{\sigma}} c_{\ve{r}+\ven+\ven' ,{\bar{\sigma}}}+ c^\dagger_{\ve{r}-\ven' ,\bar{\sigma}} c_{\ve{r} ,{\bar{\sigma}}} - h.c. \right\}.
\earray

\subsection{ Disorder by a site potential}
We work out the various sum rules in the presence of  site diagonal disorder produced by a (random) potential
$K_d= \sum_r \V{\ve{r}} \; n_{\ve{r}}$ , where we use the subscript $d$ to denote the disorder contribution to various objects. While such a potential  disorder does not change the electrical current operator, it does change the heat current, and it is easy to see that 
\barray  (\jq)_d & = & \frac{i}{2 \hbar} \sum_{\eta,\sigma} \eta_x t(\ven) (\V{\ve{r}} + \V{\ve{r}+\ven}) c^\dagger_{\ve{r}+\ven , {\sigma}} c_{\ve{r},{{\sigma}}} \nonumber \\
& = &  \frac{1}{2} \sum_{\ve{l},\sigma} \hat{V}_{- \ve{l}} (\vx{\ve{p}+\ve{l}} + \vx{\ve{p}}) c^\dagger_{\ve{p} + \ve{l} ,\sigma} c_{\ve{p},\sigma},
\earray
where we defined the fourier transform of the potential  $\hat{V}_{\ve{k}}=\frac{1}{\cal{L}} \sum_{\ve{r}} \exp( i \ve{k}.\ve{r}) V_{\ve{r}}$.
From these definitions, we   find the disorder  contribution to the thermopower sumrule
\barray
(\Phi^{xx})_d &=& \frac{q_e}{2}  \sum_{\ve{l},\ve{p},\sigma} \hat{V}_{- \ve{p}} \left( \frac{ \partial^2 \varepsilon_{\ve{l}+\ve{p}}}{\partial l_x^2} +\frac{ \partial^2 \varepsilon_{\ve{l}}}{\partial l_x^2}\right) c^\dagger_{\ve{p} + \ve{l} ,\sigma} c_{\ve{l},\sigma} \nonumber \\
&=& \frac{1}{2 \hbar} \sum_{\eta,\sigma } \eta_x^2 \; t(\ven)\; (\V{\ve{r}} + \V{\ve{r}+\ven}) c^\dagger_{\ve{r}+\ven , {\sigma}} c_{\ve{r},{{\sigma}}}
\earray
We compute the disorder contribution to the thermal conductivity sum rule as
\barray
\hbar (\Theta^{xx})_d &=&  \frac{1}{4} \sum_{\ve{r}, \ven,\sigma} t(\ven) \eta_x^2 ( V_{\ve{r}}+ V_{\ve{r}+\ven} )^2 c^\dagger_{\ve{r}+\ven ,{\sigma}} c_{\ve{r},\sigma}  \nonumber \\
 && -  \mu  \sum_{\ve{r}, \ven,\sigma} t(\ven) \; \eta_x^2 \; ( V_{\ve{r}}+ V_{\ve{r}+\ven} ) c^\dagger_{\ve{r}+\ven ,{\sigma}} c_{\ve{r},\sigma}  \nonumber \\
&& +\frac{U}{2} \sum_{\ve{r}, \ven,\sigma} t(\ven) \eta_x^2 ( V_{\ve{r}}+ V_{\ve{r}+\ven} ) ( n_{\ve{r},\bar{\sigma}}+ n_{\ve{r}+\ven ,\bar{\sigma}}) c^\dagger_{\ve{r}+\ven ,{\sigma}} c_{\ve{r},\sigma}  \nonumber \\
&&- \frac{1}{8} \sum_{\ven,\ven',\sigma} t(\ven) t(\ven') (\eta_x+\eta'_x)^2 \left\{ 3 V_{\ve{r}}+ V_{\ve{r}+\ven }+ V_{\ve{r}+\ven'}+  3 V_{\ve{r}+\ven+\ven' } \right\} c^\dagger_{\ve{r}+\ven+\ven' ,{\sigma}} c_{\ve{r},\sigma} 
\earray
 This expression can be written in fourier space as
\barray
(\Theta^{xx})_d &=&  \frac{1}{4} \sum_{\ve{l},\ve{p},\sigma} \left\{ 2 \frac{d}{d p_x} ( \xi_{\ve{p}}+ \xi_{\ve{p}+\ve{l}}) + ( v^x_{\ve{p}}+v^x_{\ve{p}+\ve{l}})^2 \right\} \; \hat{V}_{-\ve{l}} \; c^\dagger_{ \ve{p}+\ve{l},\sigma} c_{ \ve{p},\sigma} \nonumber \\
&& +  \frac{1}{4} \sum_{\ve{l},\ve{p},\ve{l'},\sigma}  \frac{d}{d p_x} \left\{  v^x_{\ve{p}}+v^x_{\ve{p}+\ve{l}} + v^x_{\ve{p}+\ve{l'}} + v^x_{\ve{p}+\ve{l}+\ve{l'}} \right\} \; \hat{V}_{-\ve{l}} \hat{V}_{-\ve{l' }}\; c^\dagger_{ \ve{p}+\ve{l}+\ve{l'},\sigma} c_{ \ve{p},\sigma} \nonumber \\
&& + \frac{U}{2 \cal{L}} \sum_{\ve{l},\ve{l'},\ve{p},\ve{q},\sigma} \frac{d^2}{d p_x^2} \left\{ \varepsilon_{\ve{p}}+\varepsilon_{\ve{p}+\ve{q}}+\varepsilon_{\ve{p}+\ve{l}}+\varepsilon_{\ve{p}+\ve{l}+\ve{q}}   \right\}\; \hat{V}_{\ve{l}} \;c^\dagger_{ \ve{q}+\ve{l'},\bar{\sigma}} c_{ \ve{l'},\bar{\sigma}} \;c^\dagger_{ \ve{p},\sigma} c_{ \ve{p}+\ve{l}+\ve{q},\sigma}  \nonumber \\
\earray
Here we need to average the final expression over the disorder potential, either perturbatively or numerically exactly. We save this task for a future work.

\subsection{Free Electron Limit and Comparison with the Boltzmann Theory}
It is easy to evaluate the various operators in the limit of $U\rightarrow 0$, and this exercise enables us to get a feel for the meaning of these various somewhat formal objects. We note that
\barray
\langle \tau^{xx} \rangle &=& {2 q_e^2} \sum_p \;\; n_{\ve{p}}\;\;   \frac{d}{d p_x} 
\left[  v^x_{\ve{p}} \right] \nonumber \\ 
\langle \Theta^{xx} \rangle &=& {2} \sum_p \;\; n_{\ve{p}}\;\; \frac{d}{d p_x} 
\left[  v^x_{\ve{p}} (\varepsilon_{\ve{p}} - \mu)^2 \right]
  \nonumber\\
\langle \Phi^{xx} \rangle &=& {2 q_e} \sum_p \;\; n_{\ve{p}}\;\; \frac{d}{d p_x} 
\left[  v^x_{\ve{p}} (\varepsilon_{\ve{p}} - \mu) \right].
\earray
Here $n_{\ve{p}}$ is the fermi function and the factor of 2 arises from spin summation. At low temperatures, we use the Sommerfield formula after integrating by parts,  and thus obtain the leading low $T$ behaviour:
\barray
\langle \tau^{xx} \rangle &=& {\cal{L}} 2 q_e^2 \rho_0(\mu) \langle (v^x_{\ve{p}})^2\rangle_{\mu} \nonumber \\
\langle \Theta^{xx} \rangle &=& {\cal{L}}T^2 \frac{2 \pi^2 k_B^2 }{3} \rho_0(\mu) \langle (v^x_{\ve{p}})^2\rangle_{\mu}
\\
\langle \Phi^{xx} \rangle &=& {\cal{L}} T^2 \frac{2 q_e \pi^2 k_B^2 }{3}\left[ 
\rho_0'(\mu) \langle (v^x_{\ve{p}})^2\rangle_{\mu} +\rho_0(\mu) \frac{d}{d \mu} \langle (v^x_{\ve{p}})^2\rangle_{\mu}
\right], \label{freetheta}
\earray
where $\rho_0(\mu)$ is the density of states per spin per site at the fermi level $\mu$ and the primes denote derivatives w.r.t. $\mu$,  the average is over the Fermi surface as usual. These formulas are indeed very close to what we expect from Boltzmann theory. Indeed  replacing $i/\omega \rightarrow \tau$ with a relaxation time $\tau$ in the leading (high frequency) terms in the general formulae reproduces  the familiar Drude-Boltzmann results\cite{ashcroft_mermin} for  the thermal conductivity, thermopower and the electrical conductivity. We may form the high frequency ratios
\barray
S^* &=& T \frac{ \pi^2 k_B^2 }{ 3 q_e} \frac{d}{d \mu} \ln \left[ \rho_0(\mu) \langle (v^x_{\ve{p}})^2\rangle_{\mu} \right] \nonumber \\
L^* &=& \frac{ \pi^2 k_B^2}{ 3 q_e^2}.
\earray
It is therefore clear that the high frequency result gives {\em the same}  Lorentz number as well as the thermopower that the Boltzmann theory gives in its simplest form. Clearly one can get more sophisticated with respect to the treatment of transport issues, such as a $k$ dependent relaxation time in the Boltzmann approach\cite{peierls,ziman},  provided the interesting Physics is in that direction. Our approach enables us to provide a more sophisticated treatment of {\em interactions and disorder}  by computing the averages of the  more complex operators given above, and this is useful for strongly correlated systems provided the transport relaxation issues are relatively benign.

\subsection{ Periodic Anderson Lattice}
We present the thermoelectric operator for the important case of the periodic  Anderson Lattice describing conduction electrons $c's$ that  hybridize with a correlated set of localized levels described by $f's$ as
\barray
K& = &  - \sum_{\sigma,\ve{\eta}} t(\ve{\eta}) c^\dagger_{ \ve{r}+ \ven} c_{ \ve{r} }
- \mu \sum_\sigma n_{\ve{r} \sigma} +( \varepsilon_0 - \mu) \sum_{1 \leq \nu \leq N_f} f^\dagger_{\vec{r},\nu} f_{\vec{r},\nu} \nonumber \\
& & + U  \sum_{\vec{r},\sigma, \nu < \nu'} n^{f}_{\vec{r},\nu} n^{f}_{\vec{r},\nu'} + \sum_{\nu,\sigma} V(\sigma,\nu)( c^\dagger_{\vec{r},\sigma} f_{\vec{r},\nu} + (h.c.)).
\earray
The charge current operator is exactly as it is in the case of the Hubbard model since the $f$ levels are localized, and using the same ideas as before to define an inhomogeneous energy $K(\vec{k}) $ and computing the familiar commutator Eq(\ref{thermopowerphi}) we find
the thermo electric operator
\beq
\Phi^{xx}= q_e \sum_{\ve{p}} \frac{\partial }{\partial p_x} \left\{   v^x_p( \ep- \mu)  \right\} c^\dagger_{\ve{p},\sigma } c_{\ve{p},\sigma} +  \frac{ q_e}{ 2 \hbar} \sum V(\sigma,\nu) \frac{ d^2 \varepsilon(\vec{k})}{d k_x^2} ( c^\dagger_{\vec{k},\sigma} f_{\vec{k}, \nu}+ f^\dagger_{\vec{k},\nu} c_{\vec{k}, \sigma} ).
\eeq
At $T=0$ the expectation of this operator should vanish (see the discussion section), whereby we obtain a formal expression for the 
chemical potential at $T=0$ from the above.

 \section{Strong Coupling Models:\\ The Heisenberg Model and the $U=\infty$ Hubbard Model}
In this section we consider the case of strong coupling, and present the Thermal and Thermoelectric operators for the
case of the Heisenberg model and the infinite correlation limit of the Hubbard model.

\subsection{Heisenberg model}
 We first  obtain the thermal operator for the ever popular Heisenberg model in any dimension described  by the Hamiltonian
\beq
H= \sum_{\ve{r}} H_{\ve{r}} \; ; \;\;\; \; H_{\ve{r}} = \frac{1}{2} \sum_{\ven} J_{\ven} \vec{S}_{\ve{r}}.\vec{S}_{\ve{r}+\ven}.
\eeq
Clearly $H(\ve{k})= \sum e^{ i \ve{k}.\ve{r}} H_{\ve{r}}$  and from the conservation law of local energy we obtain the energy current operator
\beq
\je(k_x) = \frac{1}{4} \sum_{ \ve{r}, \ven_1,\ven_2} (\eta_{1,x}- \eta_{2,x})J_{\vec{\eta}_{1}} J_{\vec{\eta}_{2}} (\ve{S}_{\ve{r}+\vec{\eta}_{1}}\times \ve{S}_{\ve{r}+ \vec{\eta}_{2}}. \ve{S}_{\ve{r}} )\; e^{i k_x x}.
\eeq 
Using the standard commutators we find the thermal operator to be a four spin operator in general. It is expressed as
\barray
\Theta^{xx} & = & \frac{1}{8} \sum_{\ve{r},\ven_1,\ven_2,\ven_3}(\eta_{1,x}- \eta_{2,x})J_{\vec{\eta}_{1}} J_{\vec{\eta}_{2}} J_{\vec{\eta}_{3}} \;\; [ \eta_{3,x}
(\vec{S}_{\vec{r}+\ven_1}\times \vec{S}_{\vec{r}+\ven_2}).(\vec{S}_{\vec{r}}\times \vec{S}_{\vec{r}+\ven_3}) 
 \nonumber \\
&& - ( 2 \eta_{2,x} + \eta_{3,x})  (\vec{S}_{\vec{r}+\ven_1}\times \vec{S}_{\vec{r}}).(\vec{S}_{\vec{r}+\ven_2}\times \vec{S}_{\vec{r}+\ven_3})
+( 2 \eta_{1,x} + \eta_{3,x})  (\vec{S}_{\vec{r}+\ven_1}\times \vec{S}_{\vec{r}+\ven_3}).(\vec{S}_{\vec{r}+\ven_2}\times \vec{S}_{\vec{r}})].
\earray
In this expression $\ve{r}, \ve{r}+\ven_1, \ve{r}+\ven_2 $ are  necessarily distinct sites, but $\ve{r}+\ven_3$ can coincide with the last two, and hence the order of the products needs to be treated carefully. For simple cubic lattices this expression simplifies a bit, but in essence it involves four spin correlations.

\subsection{ The $U\rightarrow \infty$ Hubbard model}
In this section we consider the limit of $U\rightarrow \infty$, and  consider the kinetic energy only, i.e. the $t$ part of the $t-J$ model, since this  is expected to dominate in transport properties, at least far enough from half filling and for $t>>J$. The addition of the $J$ part can be done without too much difficulty, but we ignore it here for brevity.
In this limit the fermionic commutation relations need to be modified into the Gutzwiller-Hubbard  projected operators\cite{hubbardops} 
\barray
\ch_{\ve{r},\sigma}  &=&  P_G  \; c_{\ve{r},\sigma} \; P_G   \nonumber \\
\left\{\ch_{\ve{r},\sigma}, \chd_{\ve{r'},\sigma'} \right\} &=&  \delta_{\ve{r},\ve{r'}}  \left\{ \delta_{\sigma,\sigma'} ( 1 - n_{\ve{r},\bar{\sigma}})  + ( 1 - {\delta_{\bar{\sigma},\sigma'}} ) \chd_{\ve{r},\sigma} \ch_{\ve{r},\bar{\sigma} } \right\} \nonumber\\  
&  \equiv &Y_{\sigma,\sigma'} \; \delta_{\ve{r},\ve{r'}}
\earray
The presence of the $Y$ factor is due to strong correlations, and makes the computation nontrivial. The number operator $n_{\ve{r},\sigma}$ is unaffected by the projection. We note down the expressions for the charge current and the energy current at finite wave vectors by direct computation:
\barray
\hat{K}(k) &=& - \sum_{\ve{r},\ven,\sigma} ( t(\ven)+ \mu \delta_{\ven,0}) \; e^{ i \ve{k}.(\ve{r}+\frac{1}{2} \ven)}\; \chd_{\ve{r}+\ven,\sigma} \ch_{\ve{r},\sigma} \nonumber \\
\jc(k) &=&  i \frac{q_e}{\hbar} \sum_{\ve{r},\ven,\sigma} \;\eta_x t(\ven)   \; e^{ i \ve{k}.(\ve{r}+\frac{1}{2} \ven)} \; \chd_{\ve{r}+\ven,\sigma} \ch_{\ve{r},\sigma} \nonumber \\
\jq(k)  &=& -\frac{i}{2 \hbar}\sum_{\ve{r},\ven,\ven',\sigma}\;( \eta_x + \eta'_x)  t(\ven) t(\ven')  \; e^{ i \ve{k}.(\ve{r}+\frac{1}{2} (\ven+\ven'))} \; Y_{\sigma',\sigma}(\ve{r}+\venp)\;\chd_{\ve{r}+\ven+\venp,\sigma'} \ch_{\ve{r},\sigma} 
- \frac{ \mu}{q_e} \jc(k) 
\earray

We evaluate the thermopower operator as:
\beq
\hbar \Phi^{xx}= -\frac{q_e}{2} \sum_{\ven,\vec{\eta',\sigma,\sigma'},\ve{r}} ( \eta_x+\eta'_x)^2 \; t(\ven)\; t(\venp) \; Y_{\sigma',\sigma}(\ve{r}+\ven) \; c^\dagger_{\ve{r}+\ven+\venp,\sigma'} c_{\ve{r},\sigma} - q_e \mu \sum_{\ven,\sigma} \; \eta_x^2 \; t(\ven) \; c^\dagger_{\ve{r}+\ven,\sigma} c_{\ve{r},\sigma}. \label{phi}
\eeq
Note that if we set $Y\rightarrow 1$, this expression reduces to the $U\rightarrow 0$ limit of the Hubbard result Eq(\ref{phihubbard}), as one would expect. We next compute the thermal operator (summing over all spin variables)
\barray
&&\hbar \Theta^{xx}= \mu \sum_{\ven,\vec{\eta',\sigma,\sigma'},\ve{r}} ( \eta_x+\eta'_x)^2 \; t(\ven)\; t(\venp) \; Y_{\sigma',\sigma}(\ve{r}+\ven) \; c^\dagger_{\ve{r}+\ven+\venp,\sigma'} c_{\ve{r},\sigma} + \mu^2 \sum_{\ven,\sigma} \; \eta_x^2 \; t(\ven) \; \chd_{\ve{r}+\ven,\sigma} \ch_{\ve{r},\sigma} \nonumber \\
&+ & \frac{1}{4}  \sum_{\ven,\venp,\venpp,\ve{r},\sigma,\sigma',\sigma''} ( \eta_x+\eta'_x+ \eta_x'') ( 2 \eta_x+\eta'_x+\eta_x'' )\; t(\ven)\; t(\venp) \; t(\venpp)\; Y_{\sigma'',\sigma'}(\ve{r}+\ven+\venp) \; Y_{\sigma',\sigma}(\ve{r}+\venp) \; \chd_{\ve{r}+\ven+\venp+\venpp,\sigma''} \ch_{\ve{r},\sigma} \nonumber \\
&+ & \frac{1}{4}  \sum_{\ven,\venp,\venpp,\ve{r},\sigma} ( \eta_x+\eta'_x) ( -  \eta_x+\eta'_x+\eta_x'' )\; t(\ven)\; t(\venp) \; t(\venpp)\;  \; \nonumber \\
&& \left[ \left\{\chd_{\ve{r}+\venp,\sigma} \ch_{\ve{r}+\venp+\venpp,\bar{\sigma}}   +
\chd_{\ve{r}+\venp+\venpp,\sigma} \ch_{\ve{r}+\venp,\bar{\sigma}} 
\right\}  \chd_{\ve{r}+\ven+\venp ,\bar{\sigma}} \ch_{\ve{r},\sigma}
-\left\{\chd_{\ve{r}+\venp,\bar{\sigma}} \ch_{\ve{r}+\venp+\venpp,\bar{\sigma}}   +
(h.c.)
\right\}  \chd_{\ve{r}+\ven+\venp ,{\sigma}} \ch_{\ve{r},\sigma}
\right]
\earray

\section{The Triangular Lattice Sodium Cobalt Oxide:   High temperature expansion for Thermopower}
 The Sodium Cobalt Oxide $Na_x Co O_2$ system is of great current interest;  the composition $x \sim .68$ gives a metal with a  high thermopower $\sim 100 \mu V/K$, which is further  highly magnetic field dependent\cite{terasaki,ong_2}. It is notable in that the underlying lattice is triangular, and it has been modeled by a triangular lattice $t-J$model with electron doping\cite{nco_theory}. 
In this section we give a brief  application of our  technique to this system, which yields an interesting formula for the high T limit of the thermopower $S^*$  that contains significant correction to the Heikes Mott formula that is used in the same high T limit.  
 A detailed study is in preparation and will be published separately, here we use the leading high T term to illustrate the advantage of the above formalism in tackling such a problem. We neglect contributions from the exchange part of the $t-J$ model and focus on the kinetic energy which is expected to dominate the transport contributions.
  Let us compute the thermopower $S^*$ from Eqs(\ref{phi},\ref{tau},\ref{sstar})
\beq
S^*= - \frac{\mu}{q_e T}+ \frac{ q_e \Delta}{ T \langle \tau^{xx} \rangle} \label{s1}
\eeq
where
\beq
\Delta=  -\frac{1}{2 \hbar} \sum_{\ven,\vec{\eta'},\ve{r}} ( \eta_x+\eta'_x)^2 \; t(\ven)\; t(\venp) \; \langle Y_{\sigma',\sigma}(\ve{r}+\ven) \;  c^\dagger_{\ve{r}+\ven+\venp,\sigma'} c_{\ve{r},\sigma} \rangle \label{Delta}
\eeq
This is a very useful alternate formula to the Heikes formula \cite{heikes,beni}, where the second term in Eq(\ref{s1}) is thrown out.
It interpolates very usefully between the standard formulas for low temperature as well as at high temperature.
The second term represents the ``transport'' contribution to the thermopower, whereas the first term is the thermodynamic or entropic part, which dominates at high temperature, as we shall show.  In fact for $S^*$ we can actually make a systematic expansion in powers of $\beta t$, unlike the dc counterpart, where it is not possible to make such an expansion. 

The computation of the different parts proceeds as follows: we show readily that (for the hole doped case) using translation invariance and with $n$ as the number of particles per site at high T,
\beq
\hbar \langle \tau^{xx} \rangle = 6 {\cal{L}} q_e^2 t \langle \chd_1 \ch_0 \rangle \sim 3 {\cal{L}} q_e^2 \beta t^2 n(1-n).
\eeq

The structure of the term Eq(\ref{Delta}) is most instructive. At high temperatures, for a square lattice we need to go to second order in $\beta t$ to get a contribution  with $\eta_x+\eta'_x \neq 0$, to the expectation of the hopping $\langle c^\dagger_{\ve{r}+\ven+\venp,\sigma'} c_{\ve{r},\sigma} \rangle$. For the triangular lattice, on the other hand,  we already have a contribution at  first order.  For the triangular lattice, corresponding to each nearest neighbor, there are precisely two neighbors where the third hop
is a nearest neighbor hop, and each of these gives the same factor $(\eta_x+\eta'_x)^2= 1/4$ hence at high temperatures
\beq
\Delta \sim  - \frac{3}{2 \hbar} {\cal{L}} t^2 \sum_{\sigma,\sigma'} \langle Y_{\sigma',\sigma}(\ven) \chd_{\ven+\venp,\sigma'} \ch_{\ve{0},\sigma} \rangle. 
\eeq
The spins must be the same to the leading order in $\beta t$ where we  generate a hopping term $\chd_{\ve{0},\sigma}\ch_{\ven+\venp,\sigma} $ from an expansion of $\exp(-\beta K)$, and hence a simple estimation yields
\beq
\Delta = - \frac{3}{4 \hbar} {\cal{L}} t^3 \beta n(1-n)(2-n) +O(\beta^3).
\eeq
This together with $\mu/ k_B T = \log(n/2(1-n)) + O(\beta^2 t^2)$ gives us the result for $ 0\leq n \leq 1$
\beq
S^*= \frac{k_B}{q_e} \left\{ \log[2(1-n)/n] - \beta t \frac{2-n}{2} + O(\beta^2 t^2) \right\},
\eeq
and 
\beq
S^*= - \frac{k_B}{q_e} \left\{  \log[2(n-1)/(2-n)] + \beta t \frac{n}{2} + O(\beta ^2 t^2) \right\}
\eeq
for $1 \leq n \leq 2$ using particle hole symmetry, with $q_e= - |e|$. Thus we can estimate the hopping parameter from the high temperature $T$ dependence of the thermopower.
From  the temperature dependence of the data of Terasaki et al(\cite{terasaki}) and assuming  $ S\sim S^*$one finds that $t = - 110^0K$, and with this, $S^*\sim 120 \mu V/K$,  fairly close to the observed value.
Thus the sign and rough magnitude of the hopping are as expected from other 
grounds\cite{nco_theory}. An interesting corollary of our analysis is that if the sign of hopping $t$ were reversed, i.e. if $t>0$ for the electron doped case, then the thermopower $S^*$ {\em would approach its high $T$ value from above}, and since the low $T$ behaviour is fixed to vanish linearly, it implies that $S^*$ (and hence presumably S),  must have a maximum, unlike the case $t<0$. It follows that a material with similar absolute value of hopping as $Na_xCoO_2$, but with the opposite sign of hopping, would have a greater and hence an even more exciting thermoelectric behaviour.

Hence, the nature of the high T expansion shows that the triangular lattice is exceptional in that the transport  corrections to the entropic part are much larger than for the square lattice; they are of $O(\beta t)$ for the first as opposed to    $O(\beta t)^2 $ for the latter. Also these are potentially much more sensitive to a magnetic field than the corresponding one for the square lattice since these $O(\beta t)$ terms are themselves functions of the magnetic field  $ \sim O(B^2)$.
In the degenerate limit, it seems quite possible that the field dependence of the 
transport part might be comparable to  the spin entropy contribution\cite{ong_2}.

The alternate  formula Eq(\ref{s1}, \ref{sstar}) for the thermopower  thus has the interesting property that it captures the expected low $T$  fermi liquid type behaviour as well as the high $T$ Heikes type behaviour. It should be most interesting to apply numerical techniques to evaluate this for all $T$, a calculation that seems quite possible with existing techniques, at least for small systems.

\section{Discussion}

It is curious that the recognition of the sum rule Eq(\ref{kappasumrule}) for the thermal conductivity, has lagged so far behind  Kubo's seminal paper in 1957\cite{kubo}. One of the reasons may be that most later workers  followed the method given in Sec IV  following the Kubo identity. This method does give the sum rule as shown here, but only when one dissects the Kubo identity  carefully, keeping the possibility of superconductivity in mind.  On the other hand, the many body linear response type method adopted in Sec II, runs into some discouragement, which is relieved only upon recognizing the role of the Identity I in leading to a sensible result, as  elaborated in the  discussion following Eq(\ref{kappa_final}).

 We have shown that the  sum rule for thermal conductivity involves the  thermal operator $\Theta^{xx}$ and the Seebeck coefficient involves $\Phi^{xx}$, which are formally evaluated in this work, for various models of current interest. We plan to return to a numerical evaluation of some of these in the context of strongly correlated matter in a future study. We make some remarks on the nature of the variables and the prospects for their evaluation.

One notable fact is the vanishing with $T^2$ of the expectation of $\Theta^{xx}$ and $\Phi^{xx}$ for a fermi gas, this is seen from the evaluation in Eq(\ref{freetheta}) explicitly. This leads to questions such as: What is the origin of this behaviour? Is this behaviour true in general? The answer is that these variables must vanish at low temperature in a fashion that is dictated by  the specific heat. This connection is a deep one and we explain it next.

Let us first  recognize that the specific heat (at constant chemical potential) can be written as an energy-energy correlation function:
\beq
C_{\mu}= \frac{1}{T}\lim_{k\rightarrow 0} \int_0^\beta d \tau \langle \hat{K}(k, -i \tau) \hat{K}(-k) \rangle \label{specific_heat},
\eeq 
and  the $T$ dependence of this correlator  is known in various systems in general terms. For metals and other fermi liquids, the specific heat at constant $N$, i.e.  $C_N$ is the usual measured one and differs from this by the thermodynamic relation
\beq
C_\mu=C_N+ T \frac{ (\partial S/\partial \mu)^2_T}{ (\partial N/\partial \mu)_T}, 
\eeq
where the correction term is small for a fermi liquid at low $T$   $\sim O(T^3)$. The thermal operator is  expressed as
\beq
\frac{\langle \Theta^{xx} \rangle}{\hbar T } = \frac{1}{d} C_\mu v_{eff}^2 \label{theta_spheat}
\eeq
where ``d'' is the spatial dimension, and  the above defined effective velocity $v_{eff}$ is given by
\barray
v_{eff}^2 & = & \lim_{k \rightarrow 0, t \rightarrow 0} \frac{ -d}{ k_x^2 T C_\mu} \frac{d^2}{dt^2} \int_0^\beta d \tau \langle \hat{K}(k, t -i \tau) \hat{K}(-k) \rangle \nonumber \\
&=& d \frac{ \int_0^\beta d \tau \langle \jq(0, -i \tau) \cdot \jq(0,0) \rangle} { \int_0^\beta d \tau \langle \hat{K}(0^+, -i \tau) \hat{K}(0^{-},0) \rangle}. \label{effective_velocity} 
\earray
Here we have used the relation $D_Q=0$ for generic systems to replace $\Theta^{xx}$ by the current current expectation.
The above formulae  show that the variable $\frac{\Theta^{xx}}{T}$ is best thought of as being the specific heat times the square of a velocity, much as in the kinetic theory of thermal conduction where $\kappa = \frac{1}{d} C v^2 \tau$, so that the temperature dependence comes predominantly from the specific heat and the effective velocity captures the dynamics, without being very sensitive to $T$. Of course it must be understood that $\Theta^{xx}$ does not contain the scattering rate, which disappears in the sum rule. From a computational point of view, it seems  best to use such a decomposition. At any rate, this line of argument gives us a qualitative understanding  for the rather  ``magical property'' of the $\Theta^{xx}$ operators listed,  namely the vanishing of their average  at low $T$  in the true ground state. 

 Our work also yields an interesting exact formula for the zero temperature chemical potential for  most metallic  many body  systems. Since  we expect the thermopower of metals to vanish in the ground state at all frequencies (being related to the entropy),  equating  the expectation of $\Phi^{xx}$ to zero  gives us  explicit formulae for the ground state chemical potential $\mu(0)$. This approach can be used in most cases, including hard core bose metallic systems.  In  the interesting case of the Hubbard model, from  Eq(\ref{phihubbard1}), we find  a formula for the ground state chemical potential as a ratio of expectation values of two operators
\barray
 \mu(0) & = & \frac{\cal{N}}{\cal{D}}\nonumber \\
 \cal{N} & =&  \frac{  U}{4} \sum_{} t(\ven) (\eta_x)^2  \langle ( n_{\ve{r}, \bar{\sigma}} + n_{\ve{r}+\ven, \bar{\sigma}}) ( c^\dagger_{\ve{r}+\ven,\sigma} c_{\ve{r},\sigma}+ c^\dagger_{\ve{r},\sigma} c_{\ve{r}+\ven,\sigma} )\rangle  \nonumber \\ 
&& -\frac{1}{2} \sum_{} ( \eta_x+\eta'_x)^2 t(\ven) t(\venp) \langle c^\dagger_{\ve{r}+\ven+\venp,\sigma} c_{\ve{r},\sigma}\rangle   \nonumber \\  
\cal{D}&=& {\sum_{n} \eta_x^2 t(\ven) \langle c^\dagger_{\ve{r}+\ven,\sigma} c_{\ve{r},\sigma} \rangle},\label{muformula}
\earray
where the denominator is recognized as essentially the stress tensor $\langle \tau^{xx}\rangle $. In the non interacting case, this reduces to  $\mu(0)= \varepsilon_F$, and thus generalizes this familiar relation to the interacting case.
 Similar formulas result for other models considered in this work, using the same idea.  Usually one has to resort to differentiating the ground state energy with respect to the number of particles, and this can be inaccurate in numerical studies.
Our formula is exact and a consequence of the vanishing of $S^*$ at $T=0$.  It is interesting that standard many body text books\cite{fetter,mahan} do not 
quote this type of expression, and hence it is interesting to  verify this in some cases where exact results are known from other arguments. The Hubbard model on bipartite lattices is a good a check of this result,  since the chemical potential at half filling (only) is easily found using particle hole symmetry to be $U/2$.  One can manipulate the expression Eq(\ref{muformula}) using the particle hole  symmetry,  and this result is easily reproduced.

We have presented the detailed form of the  thermal and thermo electric operators $\Theta^{xx}$ and $\Phi^{xx}$ for the Hubbard model, and also for the strongly correlated limit $U\rightarrow \infty$ above.    Results for other models not discussed here, can be readily found using the method presented.  We have also presented the operators that give the disorder contribution to these variables, these can in principle be evaluated in perturbation theory in the disorder and interaction, and it should be interesting to compute these as well in low dimensions where the effect of disorder is marked. The operators for the lattice thermal conductivity should be interesting in the context of non linear lattices, and lend themselves to numerical evaluation rather easily in the classical limit, and the quantum cases also seem to be manageable with existing computational resources.   

One set of applications concerns the effect of strong correlations on the thermo electric power factor and the figure of merit in narrow band systems.  These can be evaluated at high frequencies and such calculations should be useful guides to the role of interactions and band filling.  We provide in Section IX,  a simple  example of this formalism by computing the high temperature thermopower for the triangular lattice. This  important and currently popular  case models the physics of sodium cobalt oxide\cite{ong_2}. Interestingly the $T$ dependence of the thermopower leads to estimates of the band width that seem comparable to other estimations.
  
\begin{acknowledgments}
This work was supported by the grant 
 NSF-DMR 0408247. I am grateful to T. Giamarchi for enlightening discussions on the subtleties of the charge stiffness as well as 
 to him and A. P. Young for useful comments on  the manuscript.
\end{acknowledgments}

\appendix
\section{Electrical conductivity}
To complete the calculation we summarize the results for the electrical conductivity, which are already well known in literature\cite{plasmasumrule,giamarchi_bss,martin,ss,ferrell,scalapino}.
We note that in parallel to the calculation of the thermal conductivity, the conductivity can be expressed in terms of the correlations of the current operator $\jc$ as:
\beq
\sigma(\omega_c)=\frac{  i}{\hbar \omega_c } \invol \left[ \langle  \tau^{xx} \rangle  - i \int_0^\infty  e^{i \omega_c t'}\; dt' \langle[ \jc(t'), \jc(0)] \rangle \right]. 
\eeq
where the stress tensor
\barray
\tau^{xx}& = & \lim_{k \rightarrow 0} \frac{1}{k_x} [ \jc(k_x),\rho(-k_x)] \\
&=& \frac{q_e^2}{\hbar} \sum \eta_x^2 \; t(\ven) \; c^\dagger_{\ve{r}+\ven,\sigma} c_{\ve{r},\sigma}\;\;\;\; \mbox{or} \\
&=& \frac{q_e^2}{\hbar} \sum_{\ve{k},\sigma}\; \frac{d^2 \varepsilon_{\ve{k}}}{d k_x^2} \;  c^\dagger_{\ve{k},\sigma} c_{\ve{k},\sigma} \label{tau}
\earray
assuming a charge $q_e$ for the particles, and $n(\ve{r})$ the particle density at $\ve{r}$, and the charge fluctuation operator
$\rho(\ve{k})= q_e \sum \exp( i \ve{k}.\ve{r}) n(\ve{r})$.
 Again performing the Lehmann representation we find
\beq
\sigma(\omega_c)= \frac{  i}{\hbar \omega_c } \invol \left[ \langle \tau^{xx} \rangle   - \hbar \sum_{n,m} \frac{ p_n- p_m }{\epsilon_n - \epsilon_m + \hbar \omega_c} |\langle n|  \jc | m  \rangle|^2 \right].
\eeq 
We can use partial fractions and write
\beq
\sigma(\omega_c)= \frac{  i}{\hbar \omega_c  } D_c
+ \frac{  i \hbar}{ \cal{L}}  \sum_{n,m} \frac{ p_n- p_m }{\epsilon_m - \epsilon_n} \frac{ |\langle n|  \jc | m  \rangle|^2}{ \epsilon_n - \epsilon_m + \hbar \omega_c}.\label{sigma1}
\eeq
where the charge or Meissner stiffness is given by
\beq
 D_c = \invol \left[  \langle  \tau^{xx} \rangle  - \hbar  \sum_{n,m} \frac{ p_n- p_m }{\epsilon_m - \epsilon_n} |\langle n|  \jc | m  \rangle|^2  \right]. \label{eqdc}
\eeq 
The more familiar superfluid density $\rho_s(T)$ and plasma frequencies $\omega_{p}, \omega_{p,s}$ for the total and superconducting condensate  are  defined in terms of $D_c$ and $\langle \tau^{xx} \rangle$ by
\barray
\omega_{p,s}^2&\equiv& \frac{ 4 \pi q_e^2 \rho_s(T)}{m}  =  \frac{ 4 \pi D_c(T)}{\hbar}, \nonumber \\
\omega_p^2 &=&  \frac{ 4 \pi}{ \hbar {\cal{L}}} \langle \tau^{xx} \rangle,
\earray
so that Eq(\ref{eqdc}) is just the lattice version of the well known London decomposition of electronic density into superconducting and normal parts. The real part of the Lehmann rep is
\beq
Re \; \sigma(\omega) =  \pi  \delta(\hbar \omega) \bar{D}_c + \frac{\pi}{  \cal{L}} \left( \frac{1 - e^{ -\beta \hbar \omega}}{\omega}\right) \sum_{\epsilon_n \neq \epsilon_m} p_n  |\langle n|  \jc | m  \rangle|^2  \delta( \epsilon_m- \epsilon_n - \hbar \omega),
\eeq
where 
\beq \bar{D}_c= \invol \left[  \langle  \tau^{xx} \rangle - \hbar \sum_{\epsilon_n \neq \epsilon_m} \frac{ p_n- p_m }{\epsilon_m - \epsilon_n} |\langle n|  \jc | m  \rangle|^2  \right], \label{condeq2}
\eeq 
using  a cancellation between the terms with equal energy between the two terms. 

The lattice version of the f- sum  rule for the real part of the  conductivity (an even function of $\omega$) follows as 
\beq
\int_{0}^\infty Re \; \sigma(\omega) d\omega = \frac{ \pi}{ 2 \hbar \cal{L}} \langle  \tau^{xx} \rangle. \label{fsumrule}
\eeq

We may rewrite Eq(\ref{sigma1})  in a more compact form as
\beq
\sigma(\omega_c) = \frac{i }{ \hbar \omega_c} D_c  + \frac{1}{  {\cal L}} \int_0^\infty  dt \;e^{i \omega_c t } \int_0^\beta  d \tau \langle  \jc( -t - i \tau) \jc(0) \rangle, \label{eqsigma}.
\eeq
The difference between the two stiffnesses is
\beq
D_c - \bar{D}_c=  - \beta \hbar \sum_{\epsilon_n=\epsilon_m} p_n |\langle n | \jc | m \rangle|^2. \label{eq189}
\eeq
Thus if we know that $D_c=0$ by some independent argument, then this provides us with an alternate expression for $\bar{D}_c$\cite{integrable_charge}.

\end{document}